\documentclass[reprint,amsmath,amssymb,prl]{revtex4-1}

\usepackage{color}
\usepackage{graphicx}
\usepackage{bm}
\usepackage{amsmath,amssymb}
\usepackage{xspace}
\usepackage{braket}

\AtBeginDocument{\let\oldcontentsline\contentsline}
\newcommand{\notoccontentsline}[4]{\oldcontentsline{}{}{}{}}
\newcommand{\droptocpage}{\addtocontents{toc}{\let\protect\contentsline\protect\notoccontentsline}}
\newcommand{\incltocpage}{\addtocontents{toc}{\let\protect\contentsline\protect\oldcontentsline}}

\begin{document}
\title{Interacting Floquet polaritons}

\author{Logan W. Clark, Ningyuan Jia, Nathan Schine, Claire Baum, Alexandros Georgakopoulos, Jonathan Simon}

\affiliation{James Franck Institute and Department of Physics, University of Chicago, Chicago, IL 60637, USA}

\begin{abstract}
		Ordinarily, photons 
		do not interact with one another. However, atoms 
		can be used to mediate photonic interactions \cite{Birnbaum2005, Aoki2006, Thompson2013, Goban2015, Hammerer2010, Chang2014}, raising the prospect of forming synthetic materials \cite{Carusotto2013} and quantum information systems \cite{Raimond2001,Duan2001,Kimble2008,Saffman2010} from photons.
		One promising approach uses electromagnetically-induced transparency with highly-excited Rydberg atoms to generate strong photonic interactions \cite{Harris1997,Fleischhauer2000,Fleischhauer2005, Peyronel2012,Dudin2012a,Dudin2012b,Tiarks2014, Gorniaczyk2014}. Adding an optical cavity shapes the available modes and forms strongly-interacting 
		polaritons with enhanced light-matter coupling \cite{Guerlin2010, Jia2018b,Georgakopoulos2018}. 
		However, since every atom of the same species is identical, the atomic transitions available are only those prescribed by nature. This inflexibility severely limits their utility for mediating the formation of photonic materials in cavities, as the resonator mode spectrum is typically poorly matched to the atomic spectrum.
		Here we use Floquet engineering \cite{Silveri2017,Eckardt2017} to redesign the spectrum of Rubidium and make it compatible with the spectrum of a cavity, in order to explore strongly interacting polaritons in a customized space. 
		We show that periodically modulating the energy of an atomic level redistributes its spectral weight into lifetime-limited bands separated by multiples of the modulation frequency. 
		Simultaneously generating bands resonant with two chosen spatial modes of an optical cavity supports ``Floquet polaritons'' in both modes.
		In the presence of Rydberg dressing, we find that these polaritons interact strongly.
		Floquet polaritons thus provide a promising new path to quantum information technologies such as multimode photon-by-photon switching, as well as to ordered states of strongly-correlated photons, including crystals and topological fluids.
\end{abstract}

	\maketitle

\begin{figure}
	\centering
	\includegraphics{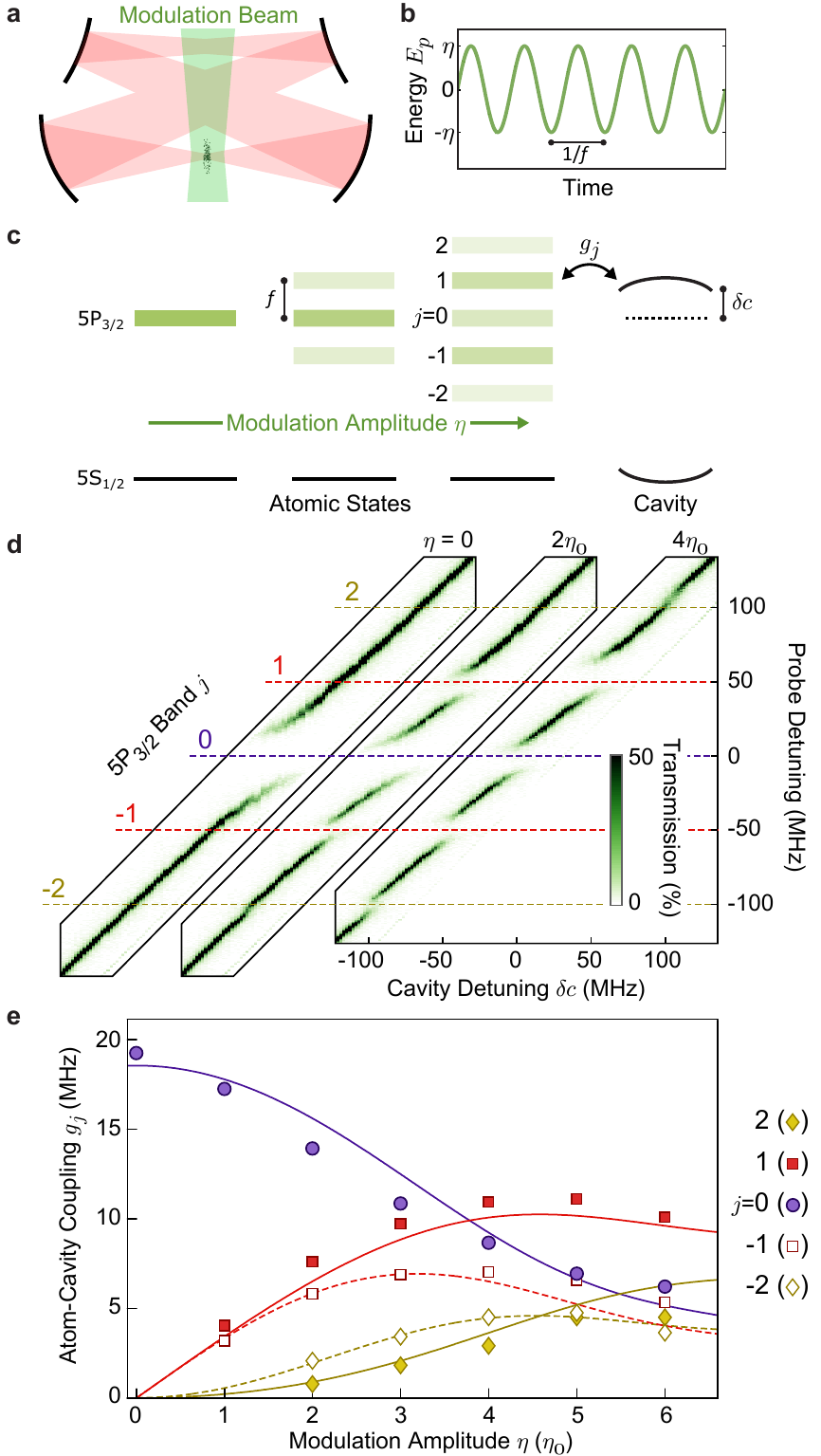} 
	\caption{\textbf{Redistributing the spectral density of atoms coupled to a cavity.} \textbf{a,} Photons in a four-mirror optical cavity are strongly coupled with excitations of a cloud of ultracold atoms. To customize the atomic spectrum, an off-resonant laser (green) sinusoidally modulates the energy $E_p$ of the $5P_{3/2}$ atomic state with amplitude $\eta$ and frequency $f$ (\textbf{b}). \textbf{c,} As the modulation amplitude increases, the $5P_{3/2}$ state progressively splits into bands at energies $E_j=jhf$ for integer $j$. Each band has a collective coupling strength $g_j$ to the cavity. \textbf{d,} Transmission spectra measured while scanning the cavity length exhibit avoided crossings, indicative of atom-cavity coupling, for up to five different bands depending on the modulation amplitude. Transmission is reported as a percentage of the value with an empty cavity. \textbf{e,} The coupling strengths $g_j$ extracted from transmission spectra are shown for the bands indicated on the right. Errorbars are smaller than the symbol heights.
	Curves (solid for positive bands, dotted for negative) show a global fit accounting for inhomogeneity and slightly asymmetric modulation (see SI.~\ref{SI:ExpBandAnalysis}). 
		\label{fig:intro}}
\end{figure}

Photons coupled to an atomic ensemble form polaritons, quasiparticles which inherit key properties from both their matter and light components \cite{Harris1997,Fleischhauer2000,Fleischhauer2005}. 
Their atomic component imbues polaritons with the interactions essential for forming non-trivial ordered phases. Their photonic component enables polaritons to propagate through space. Adding an optical cavity reshapes the space available for polaritons by defining a discrete set of modes which can couple to the atomic ensemble.
To enable the study of quantum many-body physics with polaritons \cite{Carusotto2013,Douglas2015}, exciting progress has been made in designing cavity structures to create desirable spaces for polariton propagation, including 
the realization of synthetic Landau levels for photons \cite{Schine2016,Lim2017,Schine2018}. However, the challenges of engineering these systems have thus far allowed studies of strongly interacting photons in only a single transverse mode \cite{Birnbaum2005,Aoki2006,Thompson2013,Jia2018b}. 

An intriguing approach for creating multi-mode polaritons is Floquet engineering -- the periodic modulation of parameters to generate desirable new properties in quantum mechanical systems. Floquet engineering has proven to be a powerful tool for studying quantum many-body physics with ultracold atoms \cite{Eckardt2017}, where it has enabled tests of quantum phase transitions \cite{Ma2011,Parker2013,Clark2016}, the creation of exotic new interaction processes \cite{Daley2014,Meinert2016,Clark2017}, and the development of synthetic gauge fields \cite{Lignier2007,Struck2012, Aidelsburger2013, Miyake2013,Tai2017,Clark2018} for studies of topology \cite{Jotzu2014,Aidelsburger2015,Tarnowski2017,Flaschner2018}.

Frequency modulation, the periodic variation of the energy of a quantum state, is a particularly powerful form of Floquet engineering \cite{Silveri2017}. Periodically modulating a state splits it into multiple bands at different energies, analogous to frequency modulation in signal processing \cite{Haykin2008}. Frequency modulation has enabled faster manipulation \cite{Strand2013} and efficient random access architectures \cite{Naik2017} for superconducting quantum processors via first-order sideband transitions \cite{Beaudoin2012}. In ultracold atoms, many shaken lattice experiments can be viewed as frequency modulation \cite{Eckardt2017, Silveri2017}, which has also been employed to bind diatomic molecules \cite{Thompson2005, Beaufils2009}.

In this work, we frequency modulate an atomic state to customize the coupling between atoms and photons, thereby creating strongly interacting polaritons in multiple transverse modes of a non-degenerate optical cavity. 
After loading a gas of cold atoms at the waist of the cavity, we use an intensity-modulated off-resonant laser to sinusoidally vary the energy of an excited atomic state.  This modulation splits the state into bands separated by multiples of the modulation frequency. We choose a frequency that creates bands whose energy difference matches the separation between two transverse modes of the cavity. This choice enables photons in both modes to couple with the atomic ensemble and form polaritons, which we verify by measuring the transmission spectrum. Finally, we perform a collider experiment between these Floquet polaritons, demonstrating that their strong interactions hinder multiple polaritons from entering the cavity simultaneously. We conclude by discussing the bright prospects of Floquet polaritons for many-body physics and, more broadly, customized atomic spectra for quantum science.

\begin{figure*}
	\centering
	\includegraphics{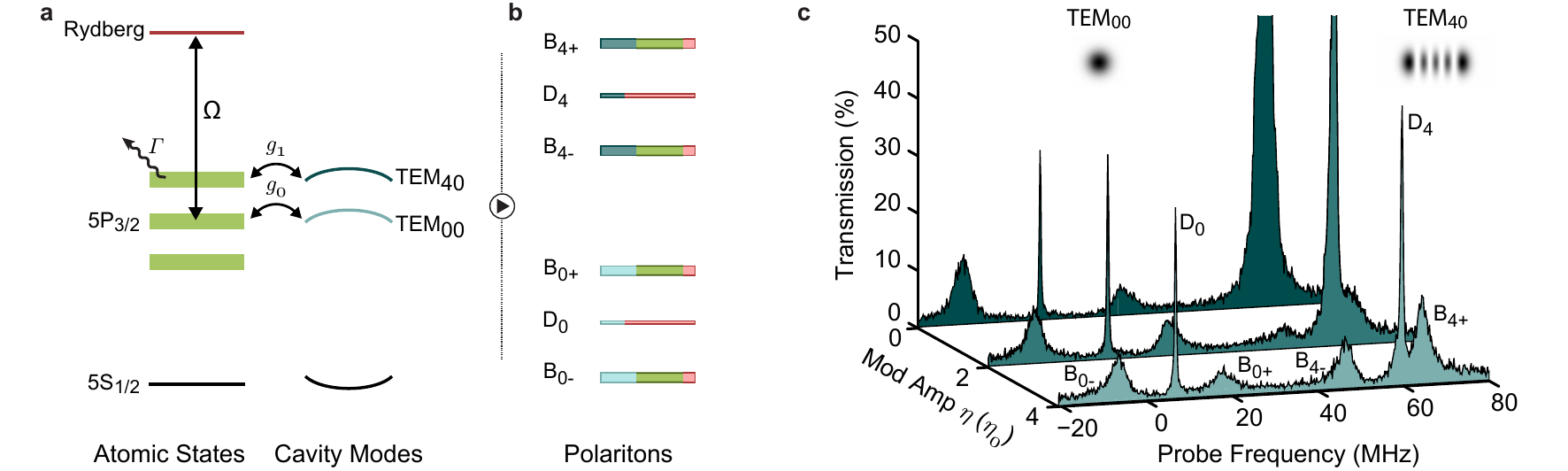} 
	\caption{\textbf{Forming Floquet polaritons in a customized space.} \textbf{a,} To form Rydberg polaritons we add a field at 480~nm to couple the $5$P$_{3/2}$ state to a Rydberg state with strength $\Omega$. When the modulation frequency $f$ matches the energy spacing between the TEM$_{00}$ and TEM$_{40}$ cavity modes, photons in both modes are resonantly coupled to excitations of the atoms with strengths determined by the resonant bands. \textbf{b,} The eigenstates of this atom-cavity system are sets of three polaritons in each spatial mode TEM$_{m0}$, consisting of two bright polaritons B$_{m\pm}$ composed primarily of a cavity photon and a collective $5$P$_{3/2}$ excitation, and one dark polariton D$_m$ composed of a cavity photon and a collective Rydberg excitation. The bright polaritons are broad due to rapid decay of their $5$P$_{3/2}$ component at rate $\Gamma=2\pi\times6$~MHz.
	\textbf{c,} With the cavity length fixed, we probe the transmission spectrum of the combined atom-cavity system. Here, we use the weakly interacting $39$S$_{1/2}$ Rydberg level to avoid blockade effects. 
	Without modulation, we observe the predicted polariton features in TEM$_{00}$, but TEM$_{40}$ exhibits only an ordinary cavity transmission line. With sufficient modulation, we detect all six polariton features predicted in panel b. Additional frequency shifts and asymmetries in the polariton spectra result from couplings to off-resonant bands (see SI.~\ref{SI:TheoryHighFreqApprox}). 
		\label{fig:EIT_00_40}}
\end{figure*}

Our experiments begin by controllably loading a sample of 300--1800 cold $^{87}$Rb atoms at the waist of a four-mirror optical cavity (Fig.~\ref{fig:intro}a). The cavity has modes near-detuned to the atomic transition between the $5$S$_{1/2}$ ground state and the $5$P$_{3/2}$ excited state at $780$~nm. In order to modulate the energy of the excited state, we expose the atoms to a multichromatic optical field tuned near the $5$P$_{3/2}\rightarrow5$D$_{5/2}$ transition at $776$~nm. This multichromatic field simultaneously cancels the constant component of the Stark shift and produces a time-varying component of the Stark shift that oscillates with tunable frequency $f$ (MHz-GHz scale) and amplitude $\eta$ (Fig.~\ref{fig:intro}b). 
We tune the modulation amplitude by adjusting the total intensity of the $776$~nm beam; amplitudes are reported in units of $\eta_0$, which corresponds to a beam intensity of approximately 5~W/mm$^2$. For more details on the experiment setup, see SI.~\ref{SI:ExpCavity}~\&~\ref{SI:ExpModScheme}.

Periodic modulation splits the excited state into bands at energies $E_j=E_0+jhf$, for integer $j$, relative to its unmodulated energy $E_0$ (Fig.~\ref{fig:intro}c). For sinusoidal modulation, the collective atom-photon coupling strengths $g_j$ at each band should take the form,
\begin{equation}
g_j\left(\eta,f\right)=g J_j\left(\frac{\eta}{hf}\right),
\label{eqn:g_bessel}
\end{equation}
where $g$ is the unmodulated coupling and $J_j$ is the $j$'th Bessel function of the first kind (see SI.~\ref{SI:TheoryBasicModulation}). 
Note that modulation does not create any new states, but rather redistributes the spectral weight of a \emph{single state} between multiple energies; this fact is reflected in the constraint that the total coupling strength $g=\sqrt{\sum_j |g_j(\eta,f)|^2}$ is unchanged (see SI.~\ref{SI:ExpModScheme}).

To test for the redistribution of the excited state into bands, we measure the transmission spectrum of a single mode of the cavity while scanning the cavity length (Fig.~\ref{fig:intro}d). Whenever the mode energy approaches the energy of an atomic band, we observe an avoided crossing in the spectrum.
Without modulation only a single avoided crossing is observed, corresponding to the original band $j=0$. With sufficient modulation additional avoided crossings become clearly visible for the first-order bands at $j=\pm1$ and the second-order bands at $j=\pm2$. The frequencies of the observed features deviate slightly from $E_j=jhf$ due to shifts from off-resonant couplings with the other bands (see  SI.~\ref{SI:TheoryHighFreqApprox}). 

We extract the coupling strengths $g_j(\eta)$ for each modulation amplitude from the widths of the avoided crossings (Fig.~\ref{fig:intro}e). While the observed coupling strengths are qualitatively similar to the prediction of Eq.~\ref{eqn:g_bessel}, the Bessel functions are distorted by inhomogeneity of the modulation beam and slight asymmetry due to higher order Stark shifts. A theoretical treatment which accounts for these two factors nicely captures the observed behavior in Fig.~\ref{fig:intro}e (see SI.~\ref{SI:ExpBandAnalysis}).

When the atomic transition is split into bands, it can couple resonantly to multiple transverse modes of the cavity simultaneously (Fig.~\ref{fig:EIT_00_40}a). In our cavity, the fundamental TEM$_{00}$ mode is conveniently close to the TEM$_{40}$ mode, which is only 52~MHz away. By modulating the atoms at a frequency near that mode separation, we simultaneously couple the TEM$_{00}$ mode with the atoms through the $j=0$ band and the TEM$_{40}$ mode with the atoms through the $j=1$ band.

At this stage, we add a $480$~nm field to couple the $5$P$_{3/2}$ state with a Rydberg state, enabling us to imbue the atomic excitations with the strong interactions of highly excited Rydberg levels. Because every band is part of the same $5$P$_{3/2}$ state, a single frequency field is sufficient to provide this Rydberg coupling. For instance, a cavity mode can couple to a $5$P$_{3/2}$ excitation through the $j=1$ band, and that same $5$P$_{3/2}$ excitation can still be subsequently coupled to a Rydberg excitation via a different band, such as $j=0$. In this work, we use a single frequency field tuned to create Rydberg coupling via band $j=0$ regardless of the resonant bands used for atom-cavity coupling. 

The eigenstates of this atom-cavity system are superpositions of collective, modulated atomic excitations with cavity photons, which we name ``Floquet polaritons'' (Fig.~\ref{fig:EIT_00_40}b). Each cavity mode TEM$_{m0}$ yields three types of polaritons \cite{Jia2016}: two bright polariton states $B_{m\pm}$ and one dark polariton state $D_m$. The bright polaritons are primarily composed of a photon in the corresponding cavity mode and a collective $5$P$_{3/2}$ excitation. Their $5$P$_{3/2}$ components with rapid decay rate $\Gamma=2\pi\times6$~MHz make them short-lived. We are primarily interested in the dark polaritons, superpositions of a cavity photon with a collective Rydberg excitation. Dark polaritons are useful for studying quantum many-body physics because they are long-lived and strongly interacting, as long as the Rydberg blockade radius is comparable to the mode size of the cavity \cite{Jia2018b}.

\begin{figure*}
	\centering
	\includegraphics{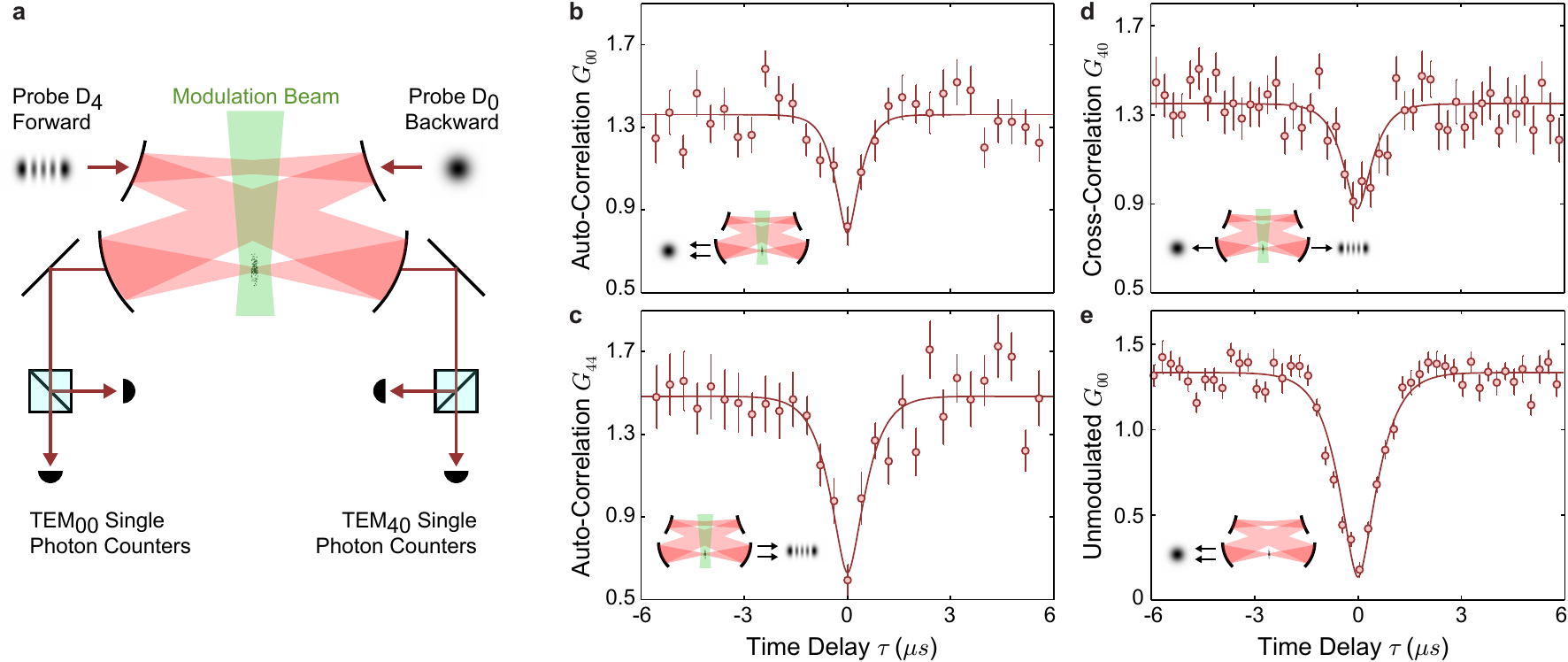} 
	\caption{\textbf{Strong interactions between Floquet dark polaritons.} \textbf{a,} We use the highly excited $100$S$_{1/2}$ Rydberg state to make dark polaritons interact strongly with each other, such that the presence of a single polariton in the cavity hinders the entry of others. To test for this blockade effect, we probe the atom-cavity system simultaneously with one laser tuned to the dark polariton D$_4$ resonance in the forward direction and a second laser tuned to the D$_0$ resonance in the backward direction. Single photon counters monitor the cavity emission in each mode. 
		With frequency modulation at $\eta=2.7~\eta_0$, photons in the TEM$_{00}$ (\textbf{b}) and TEM$_{40}$ (\textbf{c}) modes blockade further transmission in the same mode, as indicated by the correlation minima at $\tau=0$.
		\textbf{d,} Strong interactions between polaritons in different modes lead to cross-blockade. \textbf{e,} Under conditions optimized for single-mode blockade without modulation, the correlation minimum is close to zero. In panels b-e, the solid curves show fits to a generalization of the optical Bloch equations which allows one polariton to occupy either of the two modes, see SI.~\ref{SI:ExpCorrAnalysis}. Error bars indicate one standard error. 
		\label{fig:blockade}}
\end{figure*}

To detect these polaritons we measure the transmission spectrum of the atom-cavity system for a fixed cavity length (Fig.~\ref{fig:EIT_00_40}c). We choose a length which makes the TEM$_{00}$ mode resonant with the $j=0$ band and the TEM$_{40}$ mode resonant with the $j=1$ band. Without modulation, we observe the predicted polariton features in the TEM$_{00}$ mode, including two bright polaritons widely split due to strong light-matter coupling, as well as a dark polariton in the middle. The bright polaritons are broad due to the rapid decay of the $5$P$_{3/2}$ component, while the dark polariton's slow decay at rate $\Gamma_d=2\pi\times0.3$~MHz makes its transmission feature much narrower. However, without modulation there is no weight in the $j=1$ band and the atoms do not resonantly couple with the TEM$_{40}$ mode; thus, we observe a transmission feature in that mode equivalent to an empty cavity. 

Increasing the modulation amplitude couples the atoms with the TEM$_{40}$ mode by increasing the sideband strength $g_1$ at the expense of the original feature strength $g_0$. This leads to the division of the bare TEM$_{40}$ feature into the three polaritons $B_{4\pm}$ and $D_4$ as the mode becomes coupled to the atoms, while also narrowing the separation between the $B_{0\pm}$ bright polaritons due to weaker light-matter coupling in that mode. While the primary features come from the $5$P$_{3/2}$ bands that are resonantly coupled to each cavity mode, the other bands also couple with the cavity modes off-resonantly, inducing small shifts of the polariton energies. These off-resonant couplings cause the observed asymmetry between the $B_{m+}$ and $B_{m-}$ features (see SI.~\ref{SI:TheoryHighFreqApprox}).

We next perform a collider experiment between the Floquet dark polaritons (Fig.~\ref{fig:blockade}a). Here, we couple to the $100$S$_{1/2}$ Rydberg state which has a blockade radius large enough to make the polaritons interact strongly \cite{Jia2018b}. With sufficient modulation to support dark polaritons in both modes, we simultaneously probe the cavity on the $D_4$ feature in the forward direction and the $D_0$ feature in the backward direction. We then monitor the photons leaking out of the cavity in each mode, and test for photonic interactions via the correlation function $G_{mn}(\tau)$ between photons from TEM$_{m0}$ and photons from TEM$_{n0}$, separated by time $\tau$ (see SI.~\ref{SI:ExpCorrAnalysis}).

Photon antibunching in the correlation functions reveals the strong interactions between Floquet polaritons (Fig.~\ref{fig:blockade}b-d). Antibunching appears as a minimum in each correlation function at zero time delay, which indicates that the presence of just a single dark polariton in any mode of the cavity impedes the entry of a second dark polariton. We even observe cross-blockade between polaritons in different transverse modes.
In contrast, perfect coherent light without interactions, such as a laser beam, would exhibit a flat correlation function $G(\tau)=1$. Classical fluctuations, such as intensity instability, cause the correlation function to rise above one. In our system, the background correlation values result primarily from trial-to-trial fluctuations in the atom number.

Future applications of this system to quantum information, such as for multimode photon-by-photon switching, would benefit from Floquet polaritons performing at the level of polaritons optimized for blockade in a single mode (Fig.~\ref{fig:blockade}e).
Simulations using non-Hermitian perturbation theory indicate that the observed difference in performance between these two cases results from straightforwardly surmountable technical limitations (see SI~\ref{SI:ExpCorrAnalysis}). In particular, modulation splits the coupling strengths $g$ and $\Omega$ between multiple bands, weakening the resonant light-matter coupling for any particular mode. Upgrading this apparatus to achieve the atomic densities typical of free space experiments \cite{Peyronel2012} would increase $g$ by an order of magnitude, and the use of a buildup cavity for the Rydberg coupling beam with even moderate finesse would similarly enhance $\Omega$. These modifications would enable the multimode performance with frequency modulation to reach or surpass the performance shown in Fig.~\ref{fig:blockade}e, augmenting applications of this system to quantum information technology.

We have created and characterized interacting Floquet polaritons. 
These polaritons live in a completely customizable space whose modes and energetic structure (see SI.~\ref{SI:TheoryQuasienergy}) are controlled by frequency modulation of an atomic gas in an optical cavity. In particular, arbitrary control of the energies of the cavity modes is equivalent to complete control of the single-polariton dispersion.
Moreover, the structure of this space is rapidly tunable via adjustments to the modulation, raising intriguing prospects for inducing and studying polariton dynamics \cite{Zeuthen2017,Ivanov2018,Dutta2018}.
Thus, Floquet polaritons are ripe for studying strongly-correlated materials made of photons, including crystals and Laughlin states \cite{Ivanov2018,Dutta2018,Sommer2015,Ozawa2018} as well as for quantum information science \cite{Naik2017,Thompson2017}.
More broadly, this Floquet engineering scheme has a variety of other prospective applications, 
for example the matching of atomic spectra with the spectra of other physical systems for quantum information applications \cite{Saffman2010} or the tuning of spectra to enable exotic new laser cooling schemes \cite{Norcia2018}.

\droptocpage
\section{Acknowledgements}
We would like to thank Lei Feng for feedback on the manuscript. This work was supported by DOE grant DE-SC$0010267$ for apparatus construction, AFOSR grant FA9550-18-1-0317 for modeling and MURI grant FA9550-16-1-0323 for data collection and analysis. N.S. acknowledges support from the University of Chicago Grainger graduate fellowship and C.B. acknowledges support from the NSF GRFP.

\section{Author Contributions}
The experiment was designed and built by all authors. L.C., J.N. and N.S. collected the data. L.C. and J.N. analyzed the data. L.C. and J.S. developed the theory. L.C. prepared, and all authors contributed to, the manuscript. 

\section{Author Information}
The authors declare no competing financial interests. Correspondence and requests for materials should be addressed to L.C. (lwclark@uchicago.edu)

%

\setcounter{equation}{0}
\setcounter{figure}{0}
\renewcommand{\theequation}{S\arabic{equation}}
\renewcommand{\thefigure}{S\arabic{figure}}

\incltocpage
\clearpage

\tableofcontents
\appendix
\setcounter{secnumdepth}{2}

\section{Experiment}

\subsection{Experiment setup}
\label{SI:ExpCavity}
\begin{figure}
    \centering
    \includegraphics{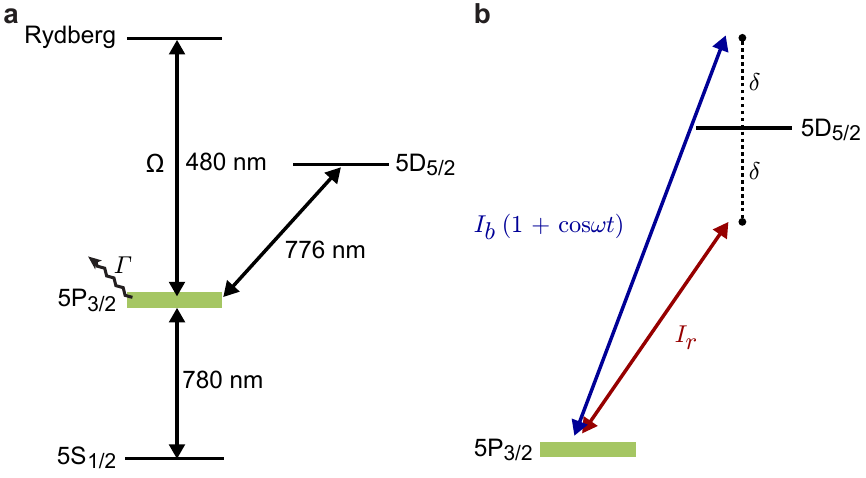}
    \caption{\textbf{Atomic level diagram.} \textbf{a,} Three key electronic transitions of $^{87}$Rb atoms enable the formation of Floquet polaritons. First, cavity photons near 780~nm couple with the $5S_{1/2}\rightarrow5P_{3/2}$ atomic transition. Second, a beam near 480~nm drives the $5P_{3/2}\rightarrow nS_{1/2}$ transition to the Rydberg level with principal quantum number $n$ at strength $\Omega$. Third, a multichromatic field near the $5P_{3/2}\rightarrow5D_{5/2}$ transition modulates the energy of the $5P_{3/2}$ state. \textbf{b,} The multichromatic field has two components with approximately opposite detunings $\mp\delta$: a red-detuned component with constant intensity $I_r$, and a blue-detuned component with sinusoidally modulated intensity $I_b(1+\cos\omega t)$. }
    \label{fig:SLevelDiagram}
\end{figure}

The experimental apparatus used for this work is described thoroughly in the supplement of Ref.~\cite{Jia2018b}. We controllably transport a sample of 300--1800 atoms $^87$Rb atoms into the region spanned by the TEM$_{40}$ mode of an optical cavity, corresponding to total atom-cavity coupling strengths of $g=$8--19~MHz on the $\ket{5S_{1/2},F=2}\rightarrow\ket{5P_{3/2},F'=3}$ atomic transition for each cavity mode. As before, a blue laser beam near $480$~nm couples the $5P_{3/2}\rightarrow nS_{1/2}$ transition for Rydberg principal quantum number $n$. A diagram of the atomic levels and transitions relevant to this work is shown in Fig.~\ref{fig:SLevelDiagram}a. As in our previous work, the atomic sample is ``sliced'' to an RMS length of approximately 10~$\mu$m which is sufficiently small for one $100$S$_{1/2}$ Rydberg excitation to blockade the formation of further Rydberg excitations. 

The cavity used in this work is the same as that described in Ref.~\cite{Jia2018b}. The cavity is non-degenerate, with a free spectral range of 2204.6~MHz and a transverse mode spacing along the vertical axis of 564.05~MHz. As a result, there is a TEM$_{40}$ mode 4*564.05~MHz-2204.6~MHz=51.6~MHz higher than the fundamental TEM$_{00}$ mode. 
Each cavity transverse mode actually contains a pair of orthogonal polarization modes, which are nearly linear. Throughout this work, we use only the modes which are approximately horizontally polarized. 

In order to ensure that the blue beam, which propagates approximately along the cavity mode axis, covers the atomic sample nearly homogeneously while still achieving sufficiently strong Rydberg coupling, we have reshaped the transverse beam profile. The blue beam now has an elliptical intensity profile with a vertical waist of $64$~$\mu$m and a horizontal waist of $20$~$\mu$m; both widths exceed the vertical (42~$\mu$m) and horizontal (12~$\mu$m) widths of the TEM$_{40}$ mode.

\subsection{Frequency modulation setup}
\label{SI:ExpModScheme}

The goal of our frequency modulation setup is to sinusoidally vary the energy $E_p$ of the $5P_{3/2}$ state. We achieve this goal using a multichromatic field near the $5P_{3/2}\rightarrow5D_{5/2}$ transition at 776~nm (Fig.~\ref{fig:SLevelDiagram}b). The field contains red-detuned (lower frequency than resonance) and blue-detuned (higher frequency) components, each approximately 1~GHz away from resonance on opposite sides of the transition, so that there is no average Stark shift of the $5$P$_{3/2}$ state. Then, we intensity modulate the blue-detuned component to cause the energy of the $5$P$_{3/2}$ state to oscillate around zero with controllable frequency $f$ and amplitude $\eta$. We tune the modulation amplitude by adjusting the total intensity of the $776$~nm laser. 

To generate the multichromatic field, we begin with a single frequency source with a total power of 20~mW which is red-detuned from the transition, locked at detuning $\delta=-\Delta$ (where $\Delta\equiv1$GHz) relative to the resonance frequency. We then generate the blue-detuned component using a fiber electro-optic modulator (EOM), driven with two RF tones at $2\Delta$ and $2\Delta+f$. This modulation generates a variety of sidebands, most of which are farther detuned from the resonance than the original beam. The sidebands which eventually produce the greatest shifts of the $5P_{3/2}$ state are the blue-detuned bands at $\delta=\Delta,\,\Delta+f$ which have similar detuning to the original beam and are first order in the RF power. These two frequency components are equivalent to a single component at $\Delta+f/2$ which is intensity modulated at frequency $f$. Therefore, our scheme can be understood simply as producing a red-detuned component with constant intensity and a blue-detuned component with modulated intensity. Finally, in order to enable large shifts of the $5P_{3/2}$ state, the entire modulated beam is sent through a tapered amplifier to achieve a maximum power of about 1~W before illuminating the atomic sample. We experimentally adjust the exact detunings and RF powers in order to achieve the largest possible $5P_{3/2}$ modulation for a given total laser power while also ensuring that the average Stark shift of the $5P_{3/2}$ state is zero. 

We have designed our setup to avoid a few detrimental side effects of frequency modulation. First, we have attempted to make the average Stark shift of the $5P_{3/2}$ state as small as possible, such that the $j=0$ band always remains at the same energy as the unmodulated $5P_{3/2}$ state. This choice is important because it ensures that the $5P_{3/2}$ band frequencies are not varying spatially due to to inhomogeneity in the modulation laser beam intensity profile, nor do they fluctuate over time due to total intensity instability. Moreover, cancelling the average Stark shift also ensures that we do not have to tune the cavity length to match a new $j=0$ band frequency every time we adjust the modulation amplitude. Note that, in this setup, the largest remaining source of temporal drift in the average Stark shift of the $5P_{3/2}$ state seems to come from the tapered amplifier, whose relative amplification of the various EOM sidebands exhibits very small fluctuations due to temperature instability. We suspect that this comes from a weak etalon effect in the amplifier chip; we observe that the $5P_{3/2}$ energy oscillates as we steadily increase the amplifier temperature. We believe that we have mitigated this effect somewhat by choosing an amplifier temperature which is at an extremum of the oscillation, making the system quadratically insensitive to temperature fluctuation around the extremum.

A second detrimental side effect comes from the off-resonant shift of the $5S_{1/2}$ state. The multichromatic field is only 4~nm detuned from the strong $5S_{1/2}\rightarrow 5P_{3/2}$ transition at $780$~nm. In principle, this shift could also be cancelled, for example by using a copropagating beam at approximately $784$~nm, but in practice cancellation was not practical in this case. We minimized the inhomogeneity of this shift, which would otherwise cause broadening of the dark polariton lines, by ensuring that beam was large (approximately round with a waist of 70~$\mu$m) compared to the horizontal cavity mode waist of 12~$\mu$m. Moreover, the $776$~nm beam propagates along the vertical axis (the long axis of the TEM$_{40}$ mode), ensuring that its intensity is approximately homogeneous across the sample along that axis without needing to increase the waist size. With sufficiently small inhomogeneity, we are able to slightly adjust the cavity length to account for the net shift of the $5S_{1/2}$ state without any deterioration of the performance in our system. 

In the future, we intend to upgrade our apparatus to use a modulation field near the $5P_{3/2}\rightarrow4D_{5/2}$ transition at 1529~nm instead of the existing 776~nm field. This substitution would have three key advantages. First, the $5P_{3/2}\rightarrow4D_{5/2}$ is stronger than the $5P_{3/2}\rightarrow5D_{5/2}$; switching transitions makes the energy modulation approximately $40\times$ larger for similar beam intensity. Second, the new detuning from the $5S_{1/2}\rightarrow 5P_{3/2}$ transition would be nearly $100\times$ greater, dramatically reducing the off-resonant shift of the $5S_{1/2}$ state. Finally, fiber EOMs operating at $1529$~nm can operate with much higher total powers, even exceeding 1~W, enabling us to put the modulator after the amplifier and thus preventing the amplifier from causing instability in the intensity ratios of the various frequency components in the field.

Throughout the text, we report modulation amplitudes in units of $\eta_0$, which is defined as the peak energy $E_p$ of the $5P_{3/2}$ state in each modulation cycle for a total $776$~nm beam intensity of approximately 5~W/mm$^2$. Based on the fit result of Fig.~1e, detailed in the next section, we estimate that $\eta_0=h\times17(1)$~MHz.

Note that the three sets of experiments reported in this work each employ a slightly different modulation frequency. In Fig.~1 we use a frequency of $f=53$~MHz, for Fig.~2 $f=58$~MHz, and for Fig.~3 $f=54$~MHz. Each of these frequencies is close to the bare cavity mode spacing of $52$~MHz between the TEM$_{00}$ and TEM$_{40}$ modes, but differs slightly to compensate for the shifts which come from coupling to off-resonant bands, see SI.~\ref{SI:TheoryHighFreqApprox}.

\subsection{Band strength analysis}
\label{SI:ExpBandAnalysis}

We extract the atom-cavity coupling strengths $g_j$ shown in Fig.~1e by independently fitting each vacuum Rabi splitting feature in the measured spectra (Fig.~1d) with the function \cite{Jia2016},
\begin{equation}
T(\delta{f}, \delta{c})=T_{0}\frac{(\kappa/2)^{2}|\tilde{\Gamma}|^{2}}{|\tilde{g}^{2}-\tilde{\kappa}\tilde{\Gamma}|^{2}}
\label{eqn:VRS_fitfunc}
\end{equation}
where $\tilde{g}$ is the atom-cavity coupling of the observed feature, the decay rate of cavity photons is $\kappa\equiv2\pi\times1.6$~MHz, the decay rate of the excited state is $\Gamma\equiv2\pi\times6$~MHz, and we define $\tilde{\kappa}\equiv-i\kappa/2+2\pi\delta{c}+\delta_{0}-2\pi\delta{f}$ and $\tilde{\Gamma}\equiv-i\Gamma/2+\delta_{e}+\delta_{0}-2\pi\delta{f}$ to account for the angular detunings of the excited state $\delta_e$ and an overall shift $\delta_0$ of the spectral feature which can arise from the off-resonant couplings (see SI.\ref{SI:TheoryModels} and SI.~\ref{SI:TheoryHighFreqApprox}). Note that the cavity detuning $\delta{c}$ and the probe detuning $\delta{f}$ are \textit{linear frequencies}, as defined in the main text, and thus require the additional factors of $2\pi$, as shown.
The features for bands $j=\pm1$ only become detectable for $\eta\geq\eta_0$, and the features for bands $j=\pm2$ only become detectable at $\eta\geq2\eta_0$.

In Fig.~1e we present a global fit to the atom-cavity coupling strengths $g_j(\eta)$ for bands $j=0,\,\pm1\,\pm2$. We simultaneously fit all 29 values of $g_j$ across all bands with a function of the form $g_j=\bar{g}A_j(\eta)$ where the total atom-cavity coupling strength $\bar{g}$ is used as the first fitting parameter. The relative amplitudes are given by,
\begin{multline}
A_j(\eta)=\Bigg[\frac{1}{2\upsilon}\int_{1-\upsilon}^{1+\upsilon}d\Upsilon\times\\
\left(\sum_{m=-2}^2 (-1)^m J_m(S\eta^2/\eta'^2\Upsilon^2)J_{j-2m}(\eta/\eta'\Upsilon)\right)^2\Bigg]^{1/2}.
\label{eqn:GlobalBandFit}
\end{multline}
The bottom line of Eq.~\ref{eqn:GlobalBandFit} is based on the theoretical prediction (Eq.~\ref{eqn:AsymmetricBandStrengths}) in SI.~\ref{SI:TheoryAsymmetricModulation}, and includes the fitting parameters $\eta'$ to provide the overall scale of the modulation amplitude and $S$ to account for strength of the serrodyne-like asymmetry in the modulation waveform. Empirically, we find that expanding the indices of the sum beyond $m=\pm2$ has negligible impact on the fit results. The rest of the function accounts for inhomogeneity in the effective modulation strength in the atomic sample. We expect inhomogeneity from two sources: first, the intensity inhomogeneity in the $776$~beam across the sample, and second, from the random Zeeman levels of the atoms in our unpolarized sample. Each randomized Zeeman level in the ground-state manifold is coupled by cavity photons to different Zeeman levels of the $5P_{3/2}$ state, which have different modulation amplitudes due to Clebsch-Gordan coefficients. To approximate the effect of these inhomogeneities, we ``blur'' the coupling strengths by taking the root-mean-square (RMS) of the atom-cavity coupling strength over a flat-top distribution of different modulation amplitudes, whose width $\upsilon$ is the fourth (and final) fitting parameter. The fit shown in Fig.~1e yielded $\bar{g}=18.7(4)$~MHz, $S=0.12(2)$, $\eta'=3.1(1)\times\eta_0$ and $\upsilon=0.6(1)$. In theory, we should have $\eta'=hf$ (see SI.~\ref{SI:TheoryAsymmetricModulation}). Therefore, using the value $f=53$~MHz from Fig.~1, the fitted value of $\eta'$ yields our estimate of $\eta_0=h\times17(1)$~MHz reported in the previous section. 

\begin{figure}
     \centering
     \includegraphics{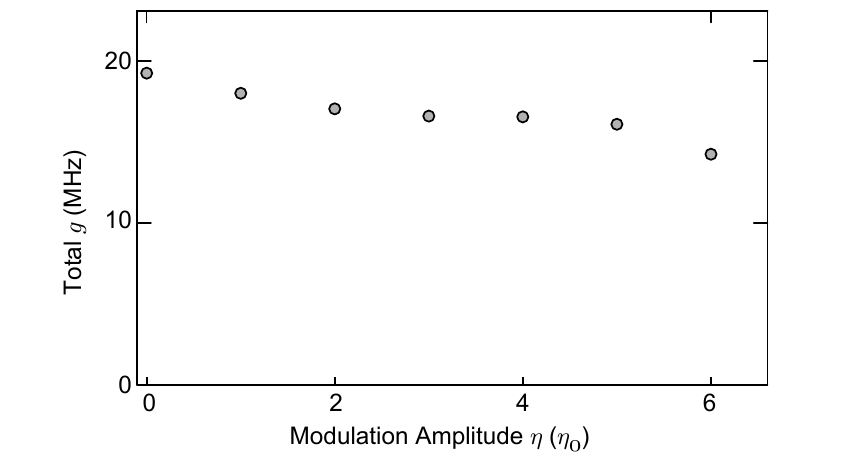}
     \caption{\textbf{Total atom-cavity coupling strength} The total atom-cavity coupling strength $g=\sqrt{\sum_j |g_j|^2}$ is nearly constant as the modulation amplitude is increased, consistent with the theoretical prediction (see SI.~\ref{SI:TheoryBasicModulation}). We attribute the slight decrease in the total $g$ to the admixture of the $5D_{5/2}$ state, which does not couple to the cavity photons, into the $5P_{3/2}$ bands.}
     \label{fig:SgTotal}
 \end{figure}

Regardless of the modulation waveform, we expect the observed band strengths to satisfy $g(\eta)=\sqrt{\sum_j |g_j(\eta)|^2}$ because we are merely redistributing the original coupling $g$ among multiple bands. For example, the predictions for both the symmetric (Eq.~\ref{eqn:g_bessel}) and asymmetric (Eq.~\ref{eqn:AsymmetricBandStrengths}) cases satisfy this constraint. The experimentally observed total coupling strengths are shown in Fig.~\ref{fig:SgTotal}. While the total strength is indeed nearly constant, it is lowered by about $25\%$ at the maximum modulation amplitude $\eta=6~\eta_0$ that we have tested. This slight decrease can be attributed to a variety of effects. First, the high intensity of the modulation beam, which leads to the nonlinear effects discussed in SI.~\ref{SI:TheoryAsymmetricModulation}, also causes a significant admixture of the $5D_{5/2}$ state into the $5P_{3/2}$ bands. Since the dipole matrix element between $5S_{1/2}$ and $5D_{5/2}$ is essentially zero, this admixture should indeed reduce the atom-cavity coupling strength. It is also possible that scattering or antitrapping caused by the 776~nm beam are reducing the number of atoms near the waist of the cavity. 

Frequency modulation of the $5P_{3/2}$ state is also expected to affect the Rydberg coupling $\Omega$, as discussed in SI.~\ref{SI:TheoryBasicModulation}. Here, the theory predicts that \emph{the effective $\Omega$ should be the same for every feature}, because regardless of the band which resonantly couples to each cavity mode, the collective Rydberg states are all degenerate and only a single frequency blue beam couples each collective $5P_{3/2}$ state to the corresponding collective Rydberg excitation. In this work, the blue beam resonantly drives the transition from the $j=0$ band to the Rydberg state, and therefore we expect every feature to exhibit Rydberg coupling strength of approximately $\Omega_0=\Omega J_0\left(\eta/hf\right)$, for symmetric modulation (SI.~\ref{SI:TheoryBasicModulation}). 

\begin{figure}
    \centering
    \includegraphics{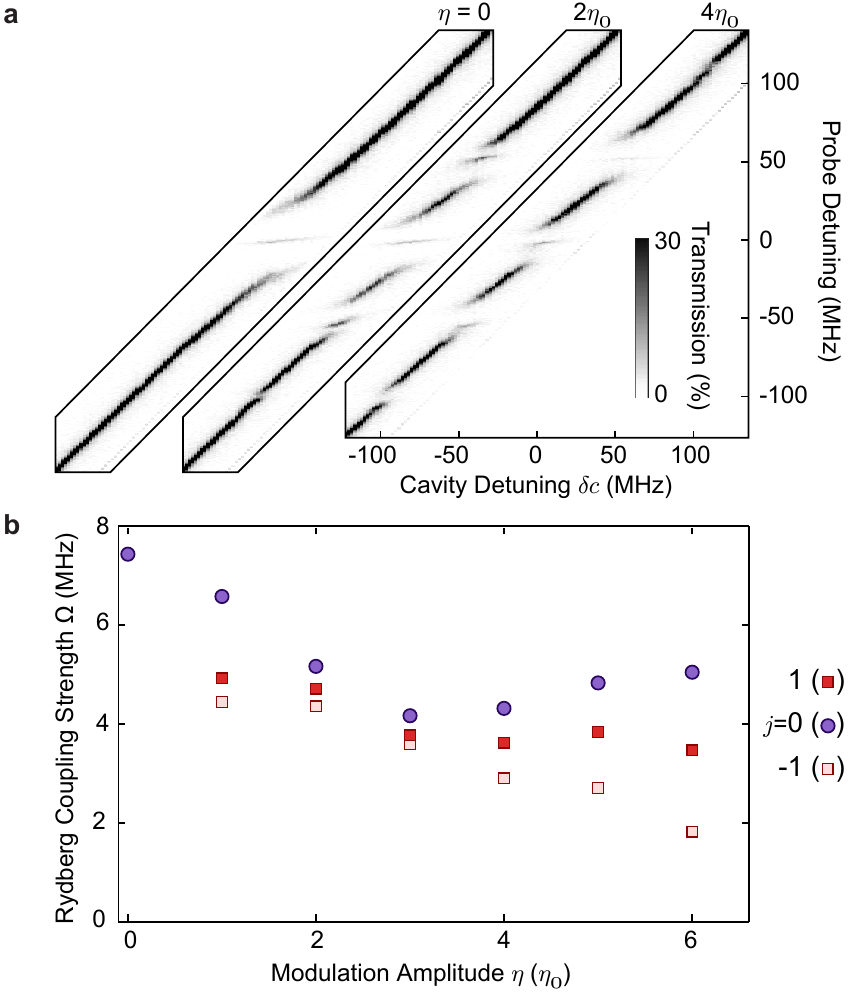}
    \caption{\textbf{Rydberg coupling strengths with frequency modulation.} \textbf{a,} Electromagnetically-induced transparency spectra observed while probing the atom-cavity system for a range of cavity lengths in the presence of the Rydberg coupling field (compare to Fig.~\ref{fig:intro}d). Even though the Rydberg coupling field contains only a single frequency, resonant with the unmodulated transition $5P_{3/2}\rightarrow39S_{1/2}$, dark polariton peaks appear at each of the $5P_{3/2}$ bands in the presence of modulation. Off-resonant shifts cause the spectra to be asymmetric (see SI.~\ref{SI:TheoryHighFreqApprox}). \textbf{b,} Rydberg coupling strengths $\Omega$ extracted from the observed spectroscopy features for the bands indicated at the right. In the idealized theory, each feature would exhibit the same $\Omega$; we attribute the deviations from this prediction to effects from the unpolarized atomic sample (see SI.~\ref{SI:ExpModScheme}).}
    \label{fig:SOmega}
\end{figure}

Similar to our treatment of the vacuum Rabi splitting features as a function of modulation amplitude in Fig.~1d-e, we present the electromagnetically-induced transparency (EIT) features as a function of modulation amplitude in Fig.~\ref{fig:SOmega}a. All of the conditions are the same as Fig.~1d-e, except that we have turned on the Rydberg coupling beam. We fit the spectroscopic features observed at each band with the function \cite{Jia2016},
\begin{equation}
T(\delta{f}, \delta{c})=T_{0}\frac{(\kappa/2)^{2}|\tilde{\Omega}^2-\tilde{\gamma}\tilde{\Gamma}|^{2}}{|\tilde{g}^{2}\tilde{\gamma}+\tilde{\kappa}(\tilde{\Omega}^2-\tilde{\gamma}\tilde{\Gamma})|^{2}},
\label{eqn:EIT_fitfunc}
\end{equation}
where $\tilde{\Omega}$ is the Rydberg coupling strength for the fitted feature, and $\tilde{\gamma}\equiv-i\gamma_R/2+\delta_R-2\pi\delta{f}$ accounts for the lifetime $\gamma_R$ and angular detuning $\delta_R$ of the Rydberg state. Note that, for these fits, $\tilde{g}$ is fixed based on the values shown in Fig.~1e, and $\gamma_R=2\pi\times1.0$~MHz is fixed to better reveal the trend of $\Omega$ with modulation amplitude $\eta$. Without constraining $\gamma_R$, there is quite a bit of spurious fluctuation in the fitted values of $\tilde{\Omega}$ which masks the overall trend. The results are not very sensitive to the exact value of $\gamma_R$ chosen: for example, increasing $\gamma_R$ by 30\% shifts the entire $\Omega$ curve for $j=0$ upward, by amounts ranging from 3\%--5\% across all $\eta$.

From fits to each feature we extract the effective Rydberg coupling strengths $\tilde{\Omega}$ for the features at each band, which are plotted in Fig.~\ref{fig:SOmega}b. For small modulation amplitudes (especially $\eta=2\sim3~\eta_0$) the observed Rydberg coupling strengths at each band are indeed quite similar, as expected. However, there are clear deviations between the coupling strengths at $\eta=\eta_0$ and $\eta\geq4\eta_0$. Note that, when $\eta=0$, there is no atom-cavity coupling on bands $j=\pm1$, so we do not observe any EIT feature and cannot extract $\tilde{\Omega}$. 

We attribute the differences in the Rydberg coupling strengths between bands to the inhomogeneity of the modulation amplitude throughout the sample. As discussed above, inhomogeneity can arise from both the $776$~nm beam profile and the unpolarized atomic sample. Qualitatively, inhomogeneity can cause $\Omega$ to deviate between bands because it causes each atom's $5P_{3/2}$ state to be distributed differently. For example, for small amplitudes such as $\eta=\eta_0$, the contributions to $g_1$ come primarily from those atoms which are being modulated more than the average, while the contributions to $g_0$ come primarily from atoms whose modulation is weaker than the average. Then, the Rydberg coupling $\Omega$ \emph{for the atoms contributing the most to $g_1$} will be smaller, because the same strong modulation which makes their contribution to $g_1$ greater makes their $\Omega$ smaller. Conversely, the coupling $\Omega$ for the atoms contributing most to $g_0$ will be higher than expected, because their modulation is weaker than the average. Thus, inhomogeneity leads to weaker $\Omega$ for the sidebands, in qualitative agreement with the trend observed in Fig.~\ref{fig:SOmega}b.

\subsection{Correlation analysis}
\label{SI:ExpCorrAnalysis}

We test for interactions between polaritons by calculating the photon-photon correlation functions,
\begin{equation}
G_{mn}(\tau)=\frac{\left< n_m(t)n'_n(t+\tau)\right>_t}{ \left<n_m\right>\left< n'_n\right>},
\label{eqn:g_definition}
\end{equation} between the photons $n_m(t)$ counted at the first detector for mode TEM$_{m0}$ and $n'_n(t)$ counted at the second detector for mode TEM$_{n0}$, where the angle brackets denote time averaging. Note that, for the cross-blockade measurement shown in Fig.~3d, we include data from all relevant combinations of the four single photon counting modules.

To obtain correlation functions representing the cavity photons, we
must account for the additional signal due to dark counts in each
detector. The measured photon counts $n_{m}^{meas}(t)$ for detector
$m$ can be written as,
\[
n_{m}^{meas}(t)=n_{m}^{cav}(t)+n_{m}^{dark}(t),
\]
where $n_{m}^{cav}(t)$ are the counts coming from cavity photons
and $n_{m}^{dark}(t)$ are the dark counts. Therefore, if we calculate
the correlations between detectors $m$ and $p$ from the measured
photon counts naively, we actually obtain,
\[
G_{mp}^{meas}(\tau)=\frac{\left<n_{m}^{meas}(t)n_{p}^{meas}(t+\tau)\right>_{t}}{\left<n_{m}^{meas}\right>\left<n_{p}^{meas}\right>}
\]
which differs from the desired correlation function of the cavity
photons,
\[
G_{mp}^{cav}(\tau)=\frac{\left<n_{m}^{cav}(t)n_{p}^{cav}(t+\tau)\right>_{t}}{\left<n_{m}^{cav}\right>\left<n_{p}^{cav}\right>}.
\]
Since dark counts occur at a constant rate for each detector (about
300~Hz in our experiments) and are uncorrelated with
the real photon counts, we calculate the correlation functions of the
cavity photons presented in the main text by using the measured counts and correcting for the average dark count rates as,
\begin{multline*}
G_{mp}^{cav}(\tau)=\frac{\left<n_{m}^{meas}(t)n_{p}^{meas}(t+\tau)\right>_{t}}{\left<n_{m}^{cav}\right>\left<n_{p}^{cav}\right>}\\
-\frac{\left<n_{p}^{dark}\right>}{\left<n_{p}^{cav}\right>}-\frac{\left\langle n_{m}^{dark}\right\rangle }{\left<n_{m}^{cav}\right>}-\frac{\left\langle n_{m}^{dark}\right\rangle \left<n_{p}^{dark}\right>}{\left<n_{m}^{cav}\right>\left<n_{p}^{cav}\right>},
\end{multline*}
where the average count rate of cavity photons $\left\langle n_{m}^{cav}\right\rangle =\left\langle n_{m}^{meas}\right\rangle -\left\langle n_{m}^{dark}\right\rangle $
can be calculated trivially. 

In our experiments, the observed correlations are affected by the strength $g$ of the atom-cavity coupling. When the number of atoms in the cavity drops too low, reducing $g$, the polaritons do not blockade each other as well. Simultaneously, the reduced $g$ causes polaritons to become more photon-like and increases the average count rate \cite{Jia2016}. In order to eliminate experimental trials in which the coupling strength is too low, we calculate the running average of the count rate across our $340\,000$ total iterations of the experiment, with an averaging range of 200~trials, and remove the iterations taken at times when the average count rate was above 6.7~kHz total across all four detectors. Only the remaining 47\% of datasets with a sufficiently low average count rate, indicating a sufficiently large $g$, were used to calculate the correlation functions shown in Fig.~3b-d. 

We find that the correlations between polaritons in our atom-cavity system are well described by a three-level model in which the cavity is either empty or contains a single polariton in one of the two modes. This model simply extends the optical Bloch equations to allow excitations in either of the two possible modes. In the limit of weak driving this model predicts correlation functions of the form \cite{WallsMilburnBook},
\begin{multline}
G_{mn}^{\mathrm{mod}}(\tau)/G^\mathrm{bkg}_{mn} = 1 + \left(1-G^\mathrm{min}_{mn}\right)\left(e^{-\gamma |\tau|}-2e^{-\gamma |\tau|/2}\right),
\label{eqn:g_model}
\end{multline}
where $G^\mathrm{bkg}_{mn}$ accounts for classical fluctuations raising the apparent correlation on long timescales, $G^\mathrm{min}_{mn}$ encodes the depth of the antibunching and accounts for imperfect blockade, and $\gamma$ is the lifetime of a dark polariton. The measured correlation functions shown in Fig.~3b-e are well explained by this model, as indicated by the fitted curves for which we used Eq.~\ref{eqn:g_model} with $G^\mathrm{bkg}_{mn}$, $G^\mathrm{min}_{mn}$, and $\gamma$ as the three fitting parameters. 

The depths of the measured correlation functions $G^\mathrm{min}_{mn}$ are consistent with our predictions using non-Hermitian perturbation theory (SI.~\ref{SI:TheoryNHPT}). From the experimental fits we extract $G^\mathrm{min}_{00}=0.58(8)$, $G^\mathrm{min}_{44}=0.42(6)$, and $G^\mathrm{min}_{40}=0.65(6)$. For comparison, our simulations predict $G_{00}^{\mathrm{sim}}(0)=0.53(1)$, $G_{44}^{\mathrm{sim}}(0)=0.44(1)$, and $G_{40}^{\mathrm{sim}}(0)=0.57(1)$. The simulations were performed with the independently characterized experimental parameters of $g=5.2$~MHz, $\Omega=1.4$~MHz, and an average, linear P-state detuning of $\delta{e}=0.6$~MHz for the TEM$_{40}$ mode, as well as $g=3.5$~MHz, $\Omega=1.0$~MHz, and $\delta{e}=1.0$~MHz for the TEM$_{00}$ mode. The agreement between experiments and simulations verifies that the observed performance is consistent with our expectations based on the current experimental parameters. 

In this work, the performance of Floquet polaritons is primarily limited by the splitting of the coupling strengths $g$ and $\Omega$ between multiple bands due to modulation. Indeed, when the atoms are not modulated, we observe much deeper antibunching $G_{00}^\mathrm{min}=0.11(2)$ as shown in Fig.~\ref{fig:blockade}e. For this optimized dataset, $g=5.5$~MHz and $\Omega=2.4$~MHz on the TEM$_{00}$ mode alone. Both of these optimized coupling strengths are less than $2.5\times$ larger than the corresponding coupling strengths achieved with the Floquet polaritons. The optimized parameters can be achieved or exceeded for the frequency modulated case with straightforward technical upgrades as discussed in the main text.


\section{Theory}

\subsection{Models for the atom-cavity system}
\label{SI:TheoryModels}

The time-dependent Hamiltonian for a multimode cavity containing a
gas of frequency-modulated three-level atoms is ($\hbar\equiv1$),

\begin{multline}
H_{0}(t)=\sum_{n}^{N_{cav}}\delta_{c}^{n}a_{n}^{\dagger}a_{n}+\delta_{e}(t)\sum_{m}^{N_{at}}\sigma_{m}^{ee}+\delta_{2}\sum_{m}^{N_{at}}\sigma_{m}^{rr}\\
+\sum_{n}^{N_{cav}}\sum_{m}^{N_{at}}(g_{mn}\sigma_{m}^{eg}a_{n}+g_{mn}^{*}\sigma_{m}^{ge}a_{n}^{\dagger})+\sum_{m}^{N_{at}}\left(\Omega_m^b\sigma_{m}^{re}+{\Omega_m^b}^{*}\sigma_{m}^{er}\right)\\
+\frac{1}{2}\sum_{n\neq m}\sigma_{m}^{rr}\sigma_{n}^{rr}U(|x_{m}-x_{n}|).
\label{eqn:HMicroscopic}
\end{multline}
This Hamiltonian is written in the rotating frame of the probe laser
with frequency $\omega_{p}$ and the blue, Rydberg coupling laser
with frequency $\omega_{b}$. There are $N_{cav}$ relevant cavity
modes and $N_{at}$ atoms in the sample. The Hamiltonian describes
the dynamics among three types of excitations: cavity photons in
mode $n$ with annihilation operator $a_{n}$ and energy $\delta_{c}^{n}\equiv i\frac{\kappa}{2}+E_{n}-\omega_{p}$
including leakage of photons from the cavity at rate $\kappa$,
excitations of atom $m$ to the 5P$_{3/2}$ state $\left|e\right\rangle _{m}$
with time-dependent energy $\delta_{e}(t)\equiv i\frac{\Gamma}{2}+E_{p}(t)-\omega_{p}$
including the decay at rate $\Gamma$, or excitations of atom $m$
to the Rydberg state $\left|r\right\rangle _{m}$ with energy $\delta_{2}\equiv i\frac{\Gamma_{r}}{2}+E_{r}-\omega_{b}-\omega_{p}$
accounting for decay at rate $\Gamma_{r}$. For convenience, we define $\sigma_{m}^{AB}\equiv\left|A\right\rangle _{m}\left\langle B\right|_{m}$.
The coupling strength between atom $m$ and cavity mode $n$ is $g_{mn}$,
which depends on the field strength $E_{n}(x_{m})$ of the cavity
mode at the two-dimensional position $x_{m}$ of the atom. The coupling from the Rydberg
coupling laser is $\Omega_m^b$, which depends on the location of atom $m$. Finally, $U(|x_{m}-x_{n}|)\equiv C_{6}/|x_{m}-x_{n}|^{6}$
encodes the interaction strength between two Rydberg atoms with the
coefficient $C_{6}$ depending strongly on the Rydberg level \cite{Saffman2010}.

Except when we are performing numerical simulations (see SI.~\ref{SI:TheoryNHPT}), we will typically work
with a simpler effective model describing the coupling of cavity modes
with collective atomic excitations \cite{Jia2018b,
Georgakopoulos2018}, 

\begin{multline}
H(t) =\sum_{n}^{N_{cav}}\delta_{c}^{n}a_{n}^{\dagger}a_{n}+\delta_{e}(t)\sum_{n}^{N_{cav}}p_{n}^{\dagger}p_{n}+\delta_{2}\sum_{n}^{N_{cav}}r_{n}^{\dagger}r_{n}\\
 +\sum_{n}^{N_{cav}}g^{n}p_{n}a_{n}^{\dagger}+h.c.+\sum_{n}^{N_{cav}}\Omega r_{n}p_{n}^{\dagger}+h.c.\\
 +\frac{1}{2}\sum_{nmpq}^{N_{cav}}U_{nmpq}r_{n}^{\dagger}r_{m}^{\dagger}r_{p}r_{q},
\label{eqn:HEffective}
\end{multline}
where we take advantage of the fact that photons in each cavity mode
$n$ couple to a unique superposition state of excited atoms with
annihilation operator $p_{n}$ for excitations in the collective $5P_{3/2}$
state and annihilation operator $r_{n}$ for excitations in the collective
Rydberg state. The many other superposition states of excited atoms which
do not couple with one of the cavity modes can be ignored for the
purposes of this supplement. We will primarily understand the behavior
of Floquet polaritons by working with this significantly compressed
Hilbert space.

\subsection{Redistributing a state using frequency modulation}
\label{SI:TheoryBasicModulation}

In this section we present the simplest theory of how frequency modulating
the $5P_{3/2}$ state redistributes its spectral density among multiple
bands, using the effective model Hamiltonian (Eq.~\ref{eqn:HEffective})
presented in the previous section. Below, in Sec.~\ref{SI:TheoryAsymmetricModulation}, we
will address our actual modulation scheme in more detail to explain
the observed asymmetry between positive and negative sidebands. 

Here, we assume that the energy of the $5P_{3/2}$ state is sinusoidally
modulated $E_{p}(t)=\bar{E}_{p}+\eta\cos(\omega t)$ with amplitude
$\eta$ and angular frequency $\omega\equiv2\pi{f}$. To put the modulated Hamiltonian
in a more useful form, we transform the collective $5P_{3/2}$ excitations,
\[
p_{n}\rightarrow\exp\left(i\frac{\eta}{\omega}\sin(\omega t)\right)p_{n}
\]
Under this transformation and using the Jacobi-Anger expansion $e^{iz\sin(\omega t)}=\sum_{n=-\infty}^{\infty}J_{n}(z)e^{in\omega t}$,
the Hamiltonian (dropping the interaction terms for simplicity) becomes,

\begin{align*}
H(t) & =\sum_{n}^{N_{cav}}\delta_{c}^{n}a_{n}^{\dagger}a_{n}+\bar{\delta}_{e}\sum_{n}^{N_{cav}}p_{n}^{\dagger}p_{n}+\delta_{2}\sum_{n}^{N_{cav}}r_{n}^{\dagger}r_{n}\\
 & +\sum_{n}^{N_{cav}}\sum_{j=-\infty}^{\infty}g_{j}^{n}e^{ij\omega t}p_{n}a_{n}^{\dagger}+h.c.\\
 & +\sum_{n}^{N_{cav}}\sum_{j=-\infty}^{\infty}\Omega_{-j}e^{-ij\omega t}r_{n}p_{n}^{\dagger}+h.c.
\end{align*}
where each collective $5P_{3/2}$ state $n$ has been redistributed
among many bands, reflected by the couplings $g_{j}^{n}\equiv g^{n}J_{j}\left(\frac{\eta}{\omega}\right)$
and $\Omega_{j}\equiv\Omega J_{-j}\left(\frac{\eta}{\omega}\right)$
for each band $j$ (recall that $\hbar\equiv1$).

While the main text emphasized the splitting of the atom-cavity coupling into bands, we see here that the Rydberg coupling $\Omega$ is also split into bands with couplings $\Omega_{j}$. However, unlike the many non-degenerate cavity modes which can couple resonantly to different bands, in our system all of the Rydberg states are degenerate. Therefore, the Rydberg coupling for each polariton mode comes from the same band $l$ determined by the frequency of the coupling field, regardless of which band $k$ is used for the atom-cavity coupling in that mode (see Fig.~\ref{fig:SOmega}). 

\subsection{Floquet polaritons in the high-frequency approximation}
\label{SI:TheoryHighFreqApprox}

In this section we use the high-frequency approximation \cite{Eckardt2017}
to derive an analytical, effective Hamiltonian for our system which
supports Floquet polaritons. As we have shown in the previous section, ignoring
Rydberg-Rydberg interactions, the periodic Hamiltonian \emph{for
each cavity mode $n$} of our modulated system can be written in
the form,
\begin{align}
H_{n}(t) & =\delta_{c}^{n}a_{n}^{\dagger}a_{n}+\bar{\delta}_{e}p_{n}^{\dagger}p_{n}+\delta_{2}r_{n}^{\dagger}r_{n}\nonumber\\
 & +\sum_{j=-\infty}^{\infty}g_{j}^{n}e^{ij\omega t}p_{n}a_{n}^{\dagger}+h.c. \label{eqn:HtforHFA}\\
 & +\sum_{j=-\infty}^{\infty}\Omega_{-j}e^{-ij\omega t}r_{n}p_{n}^{\dagger}+h.c.
\nonumber
\end{align}
In the absence of modulation, we would have $g_{0}=g$, $\Omega_{0}=\Omega,$
$g_{j\neq0}=0$, $\Omega_{j\neq0}=0$ and recover the ordinary cavity
EIT Hamiltonian. For perfect sinusoidal modulation of the P-state,
we have $g_{j}^{n}\equiv g^{n}J_{j}\left(\frac{\eta}{\omega}\right)$
and $\Omega_{-j}\equiv\Omega J_{j}\left(\frac{\eta}{\omega}\right)$
as explained in the previous section. Here, we treat the most general
form in which we allow arbitrary sideband strengths $g_{j}^{n}$ and
$\Omega_{j}$. 

The experimentally relevant case is the limit in which the cavity
mode is near-detuned to band $k$ of the 5P$_{3/2}$ state and the
Rydberg coupling laser is also near resonant for driving $5P_{3/2}\rightarrow nS_{1/2}$
on band $l$ for the chosen Rydberg principal quantum number $n$
(where $k$ and $l$ are integers). In total, these conditions allow us to write,
\begin{eqnarray}
\delta_{c}^{n}\equiv&k\omega+\epsilon_{c},\nonumber\\
\bar{\delta}_{e}\equiv&\epsilon_{p},\label{eqn:EQuasi}\\
\delta_{2}\equiv&l\omega+\epsilon_{r}\nonumber,
\end{eqnarray}
with the requirement that the ``quasienergies'' $\epsilon$ \cite{Eckardt2017} satisfy,
\[
\epsilon_{c},\epsilon_{p},\epsilon_{r}\ll\omega.
\]
We can then transform to the frame of this resonant coupling, 
\begin{eqnarray}
a_n\rightarrow e^{ik\omega t}a_n,\nonumber\\
r_n\rightarrow e^{il\omega t}r_n,\label{eqn:TransformToResonance}
\end{eqnarray}
in which the Hamiltonian becomes:

\begin{align}
H_{n}(t) & =\epsilon_{c}a_{n}^{\dagger}a_{n}+\epsilon_{e}p_{n}^{\dagger}p_{n}+\epsilon_{r}r_{n}^{\dagger}r_{n}\nonumber\\
 & +\sum_{m=-\infty}^{\infty}g_{m+k}^{n}e^{im\omega t}p_{n}a_{n}^{\dagger}+h.c.\nonumber\\
 & +\sum_{m=-\infty}^{\infty}\Omega_{-m-l}e^{-im\omega t}r_{n}p_{n}^{\dagger}+h.c.\label{eqn:HeffPerMode}
\end{align}
Since we have stipulated that the energies $\epsilon$ are all small
compared to the modulation frequency, our three states form a near-degenerate
manifold. The terms with coefficients $g_{k}$ and $\Omega_{-l}$
represent the near-resonant bands and are stationary, providing the
primary coupling which leads to the formation of polaritons. As long
as we have the additional condition that the off-resonant couplings
are small compared to the modulation frequency $g_{m\neq k},\Omega_{m\neq -l}\ll\omega$,
the rapid oscillation of the off-resonant terms enables us to use
a series approximation for the Hamiltonian. In fact, the essential physics can be understood by simply dropping all of the oscillating
terms. This is the typical rotating wave approximation, which is also
the first order high-frequency approximation. At this level one obtains
an effective Hamiltonian,
\begin{align*}
H_{F}^{(1)} & =\epsilon_{c}a_{n}^{\dagger}a_{n}+\epsilon_{e}p_{n}^{\dagger}p_{n}+\epsilon_{r}r_{n}^{\dagger}r_{n}+\\
 & g_{k}^{n}p_{n}a_{n}^{\dagger}+h.c.+\\
 & \Omega_{-l}r_{n}p_{n}^{\dagger}+h.c.
\end{align*}
whose eigenstates are polaritons with light-matter coupling strength
determined by the amplitudes of the resonant bands.

Experimentally, we often see significant shifts due to the rapidly
oscillating terms, which we can understand by applying the high-frequency
approximation beyond first order. The second order contribution to
the effective Hamiltonian $H_{F}=\sum_{m}H_{F}^{(m)}$ is,
\[
H_{F}^{(2)}=\sum_{m\neq0}\frac{H_{m}H_{-m}}{m\omega},
\]
where 
\[
H_{m}=\frac{1}{T}\int_{0}^{T}e^{-im\omega t}H_{n}(t)=H_{-m}^{\dagger}
\]
are the Fourier components of the original Hamiltonian (Eq.~\ref{eqn:HtforHFA}) and $T\equiv\frac{1}{f}$ is the Floquet period.
The total effective Hamiltonian at second order is,
\begin{align*}
H_{F} & \approx\tilde{E}_{c}a_{n}^{\dagger}a_{n}+\tilde{E}_{e}p_{n}^{\dagger}p_{n}+\tilde{E}_{r}r_{n}^{\dagger}r_{n}\\
 & +g_{k}p_{n}a_{n}^{\dagger}+h.c.+\\
 & +\Omega_{-l}r_{n}p_{n}^{\dagger}+h.c.
\end{align*}
with the quasienergies
\begin{eqnarray}
\tilde{E}_{c}=&\epsilon_{c}+\sum_{j\neq0}\frac{\left|G_{j+k}\right|^{2}}{j\omega},\nonumber\\
\tilde{E}_{p}=&\epsilon_{e}+\sum_{j\neq0}\frac{\left|\Omega_{j-l}\right|^{2}-\left|G_{j+k}\right|^{2}}{j\omega},\nonumber\\
\tilde{E}_{r}=&\epsilon_{r}+\sum_{j\neq0}\frac{-\left|\Omega_{j-l}\right|^{2}}{j\omega},\nonumber
\end{eqnarray}
shifted due to the second order terms. Note that this is the effective
Hamiltonian \emph{for each cavity mode}; since each
transverse mode of the cavity couples to independent collective atomic
excitations, we can perform this calculation for each mode considered
on its own. The only interactions between the different polariton
manifolds result from the Rydberg-Rydberg interactions, which we have
neglected in this section in order to understand the behavior of single
polaritons.

The additional shifts from the second order terms are relevant because
the expansion parameters $g_{j}/\omega$ and $\Omega_{j}/\omega$
are small but not negligible under our typical experimental conditions.
For example, the shift of a cavity mode energy $\tilde{E}_{c}$ due
to an off-resonant band with strength $g_{off}=10$ MHz would be $g_{off}^{2}/\omega=2$
MHz. These shifts have a noticeable influence on the spectrum in Fig.
2c, where the bright polariton peaks for the TEM$_{40}$ are noticeably
asymmetric. Since the TEM$_{40}$ is near-resonant with the $j=1$
band, the asymmetry is dominated by the off-resonant shift from the
srong $j=0$ band. For example, when $\eta=4\eta_{0}$, $g_{0}\approx9$
MHz, causing the TEM$_{40}$ energy to shift approximately 2 MHz toward
higher frequency, while also shifting the corresponding collective
P-state 2 MHz in the opposite direction. These shifts, as well as
shifts from the other off-resonant bands, cause the asymmetric spectra
observed in Fig. 2c.

\subsection{Quasienergy spectrum}
\label{SI:TheoryQuasienergy}

In the previous section we derived the effective Hamiltonian (Eq.~\ref{eqn:HeffPerMode}) for each cavity mode after transforming into the frame of the near-resonant band (Eq.~\ref{eqn:TransformToResonance}). When considering multiple cavity modes together, we can simply extend our transformation,
\begin{eqnarray}
a_n\rightarrow e^{ik_n\omega t}a_n,\label{eqn:TransformToResMultimode}
\end{eqnarray}
to account for the different bands $k_n$ of the $5P_{3/2}$ state which are near resonance with each cavity mode $n$. In this case, ignoring the off-resonant shifts for simplicity (that is, taking the first order High-frequency approximation), the effective Hamiltonian for the multimode system (including Rydberg interactions) is,
\begin{align}
H_F & =\sum_n^{N_\mathrm{cav}}\bigg (\epsilon_{c}^na_{n}^{\dagger}a_{n}+\epsilon_{e}p_{n}^{\dagger}p_{n}+\epsilon_{r}r_{n}^{\dagger}r_{n}\nonumber\\
 & +g_{k_n}^{n}p_{n}a_{n}^{\dagger} +\Omega_{-l}r_{n}p_{n}^{\dagger}+h.c.\label{eqn:HeffMultimode}\\
 & +\frac{1}{2}\sum_{nmpq}^{N_{cav}}U_{nmpq}r_{n}^{\dagger}r_{m}^{\dagger}r_{p}r_{q} \bigg ) \nonumber
\end{align}
As with typical cavity Rydberg polaritons, the dark polariton eigenstates of this Hamiltonian have quasienergies $E^n_\mathrm{dark}$ determined by the quasienergies of their constituent photon and Rydberg components \cite{Jia2016},
\begin{equation}
    E^n_\mathrm{D}=\epsilon_{c}^n\cos^2\theta_n + \epsilon_{r}\sin^2\theta_n
\end{equation}
where the dark-state rotation angles $\theta_n$ satisfy $\tan(\theta_n)\equiv g^n_{k_n}/\Omega_{-l}$.

Only the quasienergy appears in the effective many-polariton Hamiltonian (Eq.~\ref{eqn:HeffMultimode}), and therefore the dynamics of the Floquet system are determined by the quasienergy. That is, the specific band coupled to each cavity mode only matters to the extent that it determines the coupling strength $g^n_{k_n}$. While the coupling between the atom-cavity system and its environment, reflected for example in the spectroscopic features of Fig.~2c, may reveal ``absolute'' energies,  the many-body dynamics governed by the effective Hamiltonian depends only on the quasienergies. Therefore, even though the measured dark polariton features in Fig.~2c for $\eta=4\eta_0$ are detected at probe frequencies separated by 57.6(2)~MHz, for the purposes of dynamics within the cavity they are nearly degenerate, with quasienergies (accounting for one unit of the modulation frequency $f=58$~MHz) separated by only -0.4(2)~MHz. Moreover, this feature can be used to control the quasienergy of each polariton mode simply by varying the modulation frequency to shift the $5P_{3/2}$ bands relative to their corresponding cavity modes.

\subsection{Asymmetric band strengths}
\label{SI:TheoryAsymmetricModulation}

As detailed in SI.\ref{SI:ExpModScheme}, we modulate the energy of the $5P_{3/2}$ state by using a multichromatic driving field near the $5P_{3/2}\rightarrow5D_{5/2}$ resonance at
776 nm, see Fig.~\ref{fig:SLevelDiagram}. In general, the Stark shift $\lambda$ of the lower energy
state in a two-level atom driven with strength $\beta$ and detuning
$\delta$ is,
\[
\lambda=\frac{\mathrm{sgn}(\delta)}{2}\left(\delta^{2}+\beta^{2}\right)^{1/2}
\]
where $\mathrm{sgn}(\delta)$ is the sign of the detuning, which determines
whether the energy shift is positive or negative. 
The multichromatic field contains a blue-detuned component ($\delta>0$)
which induces a positive Stark shift of the $5P_{3/2}$ state, and
a red-detuned ($\delta<0$) component which induces a negative Stark
shift. The intensity $I_{r}$ of the red-detuned component is constant, while the intensity
$I_{b}$ of the blue-detuned component is sinusoidally modulated $I_{b}(t)={I}_{b}(1+\cos\omega t)$
where $\omega\equiv2\pi f$ is the modulation frequency which eventually
determines the splitting between the bands. The detunings of the blue
and red components from resonance are approximately equal but with
opposite sign.

\begin{figure}
    \centering
    \includegraphics[width=0.9\columnwidth]{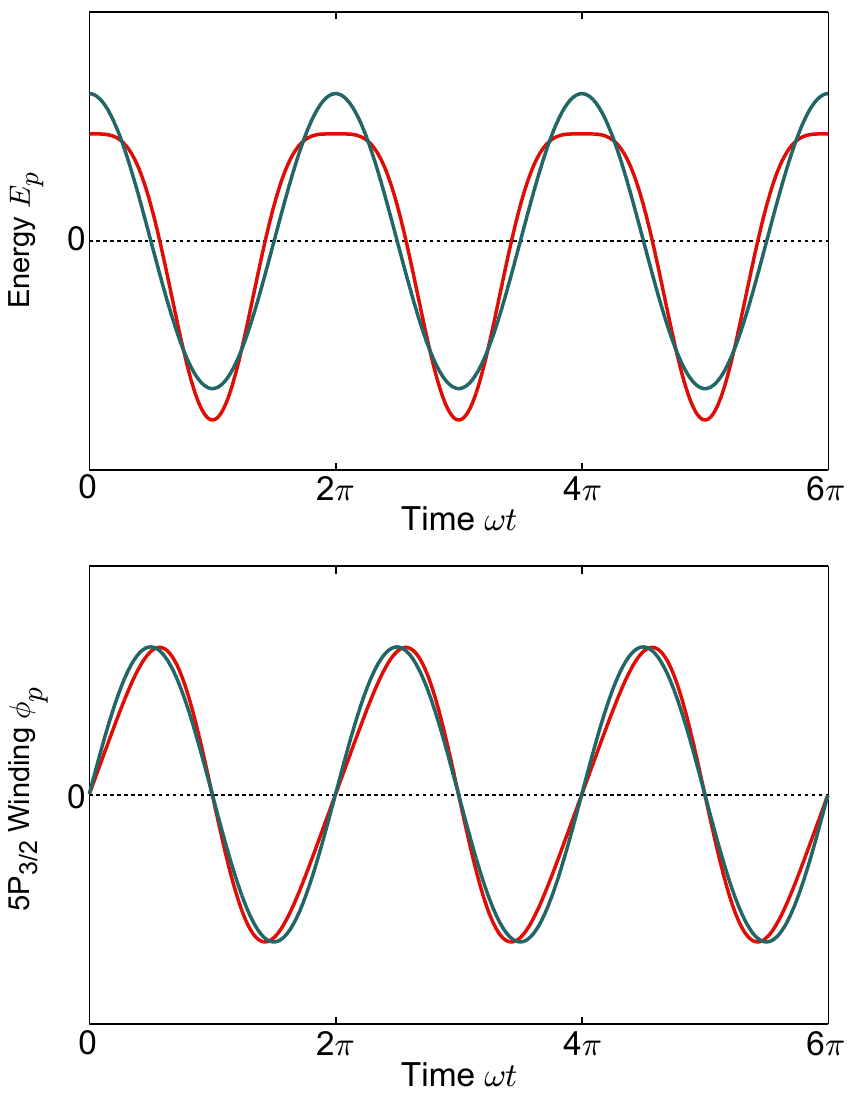}
    \caption{\textbf{Partially serrodyne modulation.} \textbf{a,} Nonlinearity in the response of the $5P_{3/2}$ energy to sinusoidal intensity modulation of the 776~nm beam causes the realistic energy waveform (red) to be asymmetric, with a sharper negative peak and flattened top peak compared to the linear response (blue). \textbf{b,} With asymmetric frequency modulation the phase winding of the state $\phi_p\propto\int E_p(t)dt$ becomes sawtooth-like (red), with shallower rises and sharper falls compared to pure sinusoidal modulation (blue). This waveform results in asymmetric sidebands even though there is no net phase winding because the average energy remains zero.}
    \label{fig:SSerrodyne}
\end{figure}

We attribute the asymmetric band strengths observed experimentally to the nonlinear response of the Stark shift to large beam intensities. 
To understand this effect, let us first consider the ideal case. Ideally, we would use a very large detuning $\left|\delta\right|\gg\left|\beta\right|$ in order to induce
a  Stark shift while keeping the amplitude of the $5D_{5/2}$ state
negligible. In this regime, to first order the Stark shift is 
\[
\lambda_{1}=\frac{\mathrm{sgn}(\delta)}{2}\left(\frac{\beta^{2}}{2\delta}\right).
\]
Since the laser intensity sets the square of the coupling strength,
$I\propto\beta^{2}$, the first order shift is directly proportional
to the intensity, $\lambda_{1}\propto I$. Therefore, in this approximation
sinusoidal modulation of the intensity would lead directly to sinusoidal
modulation of the $5P_{3/2}$ energy. 

In reality, technical limitations cause this approximation to break
down. Specifically, due to the need to avoid off-resonant Stark
shifts of the $5S_{1/2}$ state due to the $5S_{1/2}\rightarrow5P_{3/2}$
transition at 780 nm, we use detunings of only about 1 GHz. Moreover,
as can be seen from Eq. 1 of the main text, in order for the sidebands
of the modulated state to have significant weight, the amplitude of
modulation $\eta$ must be comparable to the modulation frequency
$f$. With these two constraints, we find that the second order term
in the Stark shift expansion is not negligible. Therefore, when
predicting the band strengths we must consider the second order expansion
\[
\lambda_{1}=\frac{\mathrm{sgn}(\delta)}{2}\left(\frac{\beta^{2}}{2\delta}-\frac{\beta^{4}}{4\delta^2}\right).
\]
The second order term, which is quadratic in the laser intensity,
reduces the Stark shift for large intensities. Moreover, it encodes
a Stark shift which oscillates at twice the frequency of the intensity
modulation. 

In this second order approximation, and assuming that the average
intensities of the red and blue components have been chosen to cancel
the time-averaged Stark shift, the energy of the $5P_{3/2}$ state
takes the form,
\[
E_{p}(t)=\bar{E}_{p}+x'I\cos(\omega t)-y''I^{2}\left(\cos^{2}(\omega t)-\frac{1}{2}\right)
\]
where we have absorbed numerical constants and parameters other than
the total intensity $I$ of 776 nm light into the positive coefficients
$x',y''>0$ for simplicity. This waveform ``spikes'' toward negative
energies such that the modulation is asymmetric, see Fig.~\ref{fig:SSerrodyne}a. We can equivalently write
this as,
\[
E_{p}(t)=\bar{E}_{p}+x'I\cos(\omega t)-y'I^{2}\cos(2\omega t)
\]
where $y'\equiv y''/2$. Note that, because we are considering a specific
form of nonlinearity, the relative phases of the two sinusoidal components
(including the positive signs of the coefficients) are fixed and physically
motivated; we will later see that this phase relationship results
in asymmetric sideband strengths. 

Similar to the simpler case in SI.~\ref{SI:TheoryBasicModulation}, we now make the
transformation,
\[
p_{n}\rightarrow\exp\left(ixI\sin(\omega t)-iyI^{2}\sin(2\omega t)\right)p_{n},
\]
where $x\equiv x'/\omega$ and $y\equiv y'/\omega$. Here, we transform to a frame in which amplitude in the $5P_{3/2}$ state has a phase winding over time as $\phi_p\equiv xI\sin(\omega t)-yI^2\sin(2\omega t)$, see Fig.~\ref{fig:SSerrodyne}b. The spikes in the energy waveform have caused the phase winding to have a sawtooth-like character, characteristic of serrodyne modulation \cite{Haykin2008}, which leads to highly asymmetric band strengths.

With this transformation we find that the atom-light couplings
take the form (applying the Jacobi-Anger expansion twice), 
\begin{eqnarray*}
g_{j}^{n}=g^{n}A_{j}(I),\\
\Omega_{-j}=\Omega A_{j}(I)
\end{eqnarray*}
where the strength of each band is determined by the sum,
\begin{equation}
A_{j}(I)\equiv\sum_{m=-\infty}^{\infty}(-1)^{m}J_{m}(yI^{2})J_{j-2m}(xI).
\label{eqn:AsymmetricBandStrengths}
\end{equation}
To understand this result, let us start with two simplified cases.
First, if we have no modulation of any kind ($I=0$), then all of
the amplitude is in the original band ($A_{0}=1$, $A_{j\neq0}=0$).
Second, let us consider the case where the nonlinearity is negligible
($y$=0). In this case, we have $A_{j}(I)=J_{j}(xI)$ and we have
recovered the idealized relative band strengths from SI.~\ref{SI:TheoryBasicModulation}.
In that case, since $J_{j}=(-1)^{j}J_{-j}$, the strengths of positive
and negative sidebands are equal.

In the realistic case, since we are effectively modulating at two
frequencies due to the nonlinear response, the strength of each sideband
of the 5P$_{3/2}$ state is actually determined by the sum (Eq.~\ref{eqn:AsymmetricBandStrengths}) over all possible combinations of sidebands for
the fundamental frequency and the second harmonic which add up to
the same total frequency. These different ``paths'' can interfere,
resulting in asymmetry between positive and negative sideband strengths.
As a simple example for seeing this interference, consider the case
of small but not negligible second harmonic amplitude $yI^{2}\ll1$,
where we only need to include the terms $m=\pm1,0$ (Eq.~\ref{eqn:AsymmetricBandStrengths})
The strengths $A_{\pm1}$ can then be written (also neglecting terms with typically small factors of $J_3(xI)$),
\begin{eqnarray*}
A_{1}(I)=J_{0}(yI^{2})J_{1}(xI)+J_{1}(yI^{2})J_{1}(xI),\\
A_{-1}(I)=-J_{0}(yI^{2})J_{1}(xI)+J_{1}(yI^{2})J_{1}(xI).
\end{eqnarray*}
Since these equations have different interference phases, the strengths
will generally have different magnitude $|A_{1}(I)|\neq|A_{-1}(I)|$.
Similar considerations apply for all sidebands and for any size of
the second harmonic amplitude $y.$

\subsection{Multimode non-Hermitian perturbation theory}
\label{SI:TheoryNHPT}

In this section we present a treatment of our system via non-Hermitian perturbation theory, extended from that in Ref.~\cite{Jia2018b} to include multiple transverse modes of the cavity. We use this theory to predict the correlation minima $G_{nm}(0)$ for comparison to the results shown in Fig.~3 of the main text.

We begin with the complete microscopic Hamiltonian which treats each atom independently (Eq.~\ref{eqn:HMicroscopic}). As noted in Sec.~\ref{SI:TheoryQuasienergy}, only the quasienergy of the cavity modes should be relevant to the behavior of this system. Therefore, we can completely account for the effects of Floquet engineering by making two changes to the Hamiltonian. First, we use the quasienergy (Eq.\ref{eqn:EQuasi}) of each cavity mode and Rydberg level rather than their absolute energies. Second, we account for the $5P_{3/2}$ band which is near-resonant to each the cavity mode $n$ by modifying the corresponding coupling strengths $g_{mn}$. With those modifications, the microscopic Hamiltonian can be written, 

\begin{multline}
H_{0}=\sum_{n}^{N_{cav}}\epsilon_{c}^{n}a_{n}^{\dagger}a_{n}+\epsilon_{e}\sum_{m}^{N_{at}}\sigma_{m}^{ee}+\epsilon_r\sum_{m}^{N_{at}}\sigma_{m}^{rr}\\
+\sum_{n}^{N_{cav}}\sum_{m}^{N_{at}}(g_{mn}\sigma_{m}^{eg}a_{n}+g_{mn}^{*}\sigma_{m}^{ge}a_{n}^{\dagger})+\sum_{m}^{N_{at}}\left(\Omega_m^b\sigma_{m}^{re}+{\Omega_m^b}^{*}\sigma_{m}^{er}\right)\\
+\frac{1}{2}\sum_{p\neq q}\sigma_{q}^{rr}\sigma_{p}^{rr}U(|x_{m}-x_{n}|).
\label{eqn:HMicroscopicNHPT}
\end{multline}

This Hamiltonian is perturbed by the cavity probe,
\[
V=\Omega_{p}(c^{\dagger}+c),
\]
where we allow the input mode to be an arbitrary superposition
$c$ of the cavity modes,
\[
c=\sum_{n}^{N_{cav}}C_{n}a_{n},
\]
\[
\sum_{n}|C_{n}|^{2}=1.
\]
As in typical perturbation theory we can rewrite the exact eigenstate
as a series expansion in the perturbation strength,
\[
\left|\psi_{exact}\right\rangle =\sum_{n}\left(\Omega_{p}\right)^{n}\left|\psi_{n}\right\rangle 
\]
starting with the vacuum state,
\[
\left|\psi_{0}\right\rangle =\left|0\right\rangle ,
\]
which contains no excitations at all. Noting that we have included
the effects of the rotating frame in the original Hamiltonian above,
non-Hermitian perturbation theory shows that the first-order equation
is,
\[
H_{0}\left|\psi_{1}\right\rangle =\tilde{V}\left|\psi_{0}\right\rangle 
\]
where 
\[
\tilde{V}\equiv\frac{V}{\Omega_{p}}=c^{\dagger}+c.
\]
To predict the photon correlations we must continue to the second
order equation, 
\[
H_{0}\left|\psi_{2}\right\rangle =\tilde{V}\left|\psi_{1}\right\rangle 
\]
which provides the lowest order behavior in the two excitation manifold.

\subsubsection{First order: resonator transmission spectrum}

At first order starting from the vacuum state, at most one excitation
can be in the system. Therefore, we expand the first order state
as
\[
\left|\psi_{1}\right\rangle =\left(\sum_{n}^{N_{cav}}A_{c}^{n}a_{n}^{\dagger}+\sum_{m}^{N_{at}}A_{e}^{m}\sigma_{m}^{eg}+A_{r}^{m}\sigma_{m}^{rg}\right)\left|0\right\rangle .
\]
Using the first-order non-Hermitian perturbation theory prediction we obtain the following
set of equations. For each cavity mode $n$ we have one equation,

\[
C_{n}^{*}=\epsilon_{c}^{n}A_{c}^{n}+\sum_{m}^{N_{at}}g_{mn}^{*}A_{e}^{m},
\]
and for each atom we have two equations,
\[
0=\sum_{n}^{N_{cav}}g_{mn}A_{c}^{n}+\epsilon_{e}A_{e}^{m}+(\Omega_m^b)^{*}A_{r}^{m},
\]

\[
0=\epsilon_{r}A_{r}^{m}+\Omega_m^b A_{e}^{m}.
\]
Solving the equations for the atoms yields,

\[
A_{r}^{m}=-\frac{\Omega_m^b}{\epsilon_{r}}A_{e}^{m},
\]

\[
A_{e}^{m}=\frac{1}{\frac{\left|\Omega_m^b\right|^{2}}{\epsilon_{r}}-\epsilon_{e}}\sum_{n}^{N_{cav}}g_{mn}A_{c}^{n},
\]
which we can substitute into the cavity equations to obtain,

\[
C_{n}^{*}=\epsilon_{c}^{n}A_{c}^{n}+\sum_{m}^{N_{at}}\frac{\left|g_{mn}\right|^{2}}{\frac{\left|\Omega_m^b\right|^{2}}{\epsilon_{r}}-\epsilon_{e}}A_{c}^{n}+\sum_{k\neq n}^{N_{cav}}\sum_{m}^{N_{at}}\frac{g_{mn}^{*}g_{mk}}{\frac{\left|\Omega_m^b\right|^{2}}{\epsilon_{r}}-\epsilon_{e}}A_{c}^{k}.
\]
At this stage, one can perform a straightforward inversion of a matrix of size
$N_{cav}\times N_{cav}$ to obtain a complete first-order solution. Alternatively,
we can recognize that the second term on the right-hand side represents
the strong superradiant coupling of the excited atoms to the cavity
mode, while the third term on the right-hand side repesents much weaker
subradiant emission of collective
excited states corresponding to modes $k$ into the ``wrong'' cavity
mode $n$. In the experimentally relevant limit of large atom numbers
$N_{at}\gg1$ and a homogeneous distribution of atoms, the second term dominates and the third term can be neglected, yielding the simple solution,
\[
A_{c}^{n}=\frac{C_{n}^{*}}{\epsilon_{c}^{n}+\sum_{m}^{N_{at}}\frac{\left|g_{mn}\right|^{2}}{\frac{\left|\Omega_m^b\right|^{2}}{\epsilon_{r}}-\epsilon_{e}}},
\]
which matches the original single-mode
solution (Ref.~\cite{Jia2018b}) for
each of the modes $n$ independently, while also accounting for the relative probe amplitude $C_{n}$ in each.

\subsubsection{Second order: photon-photon correlations}

Similar to our approach at first order, we can now write the second-order
solution as an expansion in the two-excitation basis,

\begin{multline*}
\left|\psi_{2}\right\rangle =\bigg(\sum_{n}^{N_{cav}}\frac{B_{cc}^{nn}}{\sqrt{2}}\left(a_{n}^{\dagger}\right)^{2}+\sum_{p>n}^{N_{cav}}B_{cc}^{np}a_{n}^{\dagger}a_{p}^{\dagger}\\
+\sum_{n}^{N_{cav}}\sum_{j}^{N_{at}}B_{ce}^{nj}a_{n}^{\dagger}\sigma_{j}^{eg}+B_{cr}^{nj}a_{n}^{\dagger}\sigma_{j}^{rg}+\sum_{j\neq k}^{N_{at}}B_{er}^{jk}\sigma_{j}^{eg}\sigma_{k}^{rg}\\
+\sum_{j>k}^{N_{at}}B_{rr}^{jk}\sigma_{j}^{rg}\sigma_{k}^{rg}+B_{ee}^{jk}\sigma_{j}^{eg}\sigma_{k}^{eg}\bigg)\left|0\right\rangle.
\end{multline*}
Matching coefficients for each kind of double-excitation, the second
order equations of motion based on the Non-Hermitian Perturbation
Theory are,
\\

\noindent for doubly occupied cavity modes $\frac{(a_{k}^{\dagger})^{2}}{\sqrt{2}}$:
\[
2\epsilon_{c}^{k}B_{cc}^{kk}+\sqrt{2}\sum_{m}^{N_{at}}g_{mk}^{*}B_{ce}^{km}=\sqrt{2}C_{k}^{*}A_{c}^{k}
\]
two photons in different modes $a_{k}^{\dagger}a_{n}^{\dagger}$:
\[
(\epsilon_{c}^{n}+\epsilon_{c}^{k})B_{cc}^{nk}+\sum_{m}^{N_{at}}g_{mn}^{*}B_{ce}^{km}+\sum_{m}^{N_{at}}g_{mk}^{*}B_{ce}^{nm}=C_{k}^{*}A_{c}^{n}+C_{n}^{*}A_{c}^{k}
\]
one photon one p-state excitation $a_{k}^{\dagger}\sigma_{m}^{eg}$:
\begin{multline*}
\epsilon_{c}^{k}B_{ce}^{km}+\sqrt{2}g_{mk}B_{cc}^{kk}+\sum_{p\neq k}^{N_{cav}}g_{mp}B_{cc}^{kp}+\sum_{n\neq m}^{N_{at}}g_{nk}^{*}B_{ee}^{nm}\\
+\epsilon_{e}B_{ce}^{km}+\left(\Omega_{m}^{b}\right)^{*}B_{cr}^{km}=C_{k}^{*}A_{e}^{m}
\end{multline*}
one photon one rydberg \textbf{ $a_{k}^{\dagger}\sigma_{m}^{rg}$}:
\[
\left(\epsilon_{c}^{k}+\epsilon_{r}\right)B_{cr}^{km}+\sum_{n\neq m}^{N_{at}}g_{kn}^{*}B_{er}^{nm}+\Omega_{m}^{b}B_{ce}^{km}=C_{k}^{*}A_{r}^{m}
\]
two p-state excitations $\sigma_{j}^{eg}\sigma_{k}^{eg}$:
\[
\sum_{n}^{N_{cav}}(g_{jn}B_{ce}^{nk}+g_{kn}B_{ce}^{nj})+2\epsilon_{e}B_{ee}^{jk}+\left(\Omega_{k}^{b}\right)^{*}B_{er}^{jk}+\left(\Omega_{j}^{b}\right)^{*}B_{er}^{kj}=0
\]
one p-state excitation, one Rydberg excitation \textbf{ $\sigma_{j}^{eg}\sigma_{k}^{rg}$}:
\[
(\epsilon_{e}+\epsilon_{r})B_{er}^{jk}+\sum_{n}^{N_{cav}}g_{jn}B_{cr}^{nk}+\Omega_{k}^{b}B_{ee}^{jk}+\left(\Omega_{j}^{b}\right)^{*}B_{rr}^{jk}=0
\]
two Rydberg excitations \textbf{ $\sigma_{j}^{rg}\sigma_{k}^{rg}$}:
\[
(2\epsilon_{r}+U(|x_{j}-x_{k}|))B_{rr}^{jk}+\Omega_{j}^{b}B_{er}^{jk}+\Omega_{k}^{b}B_{er}^{kj}=0
\]
\\

As in Ref.~\cite{Jia2018b}, we can
begin algebraically by taking advantage of the lack of direct couplings
which exchange excitations between atoms in order to reduce the $O(N^{2})$
equations here to only $O(N)$ equations which must then be solved
numerically. In particular, we essentially have $N_{at}(N_{at}-1)$
independent sets of four coupled equations, one set for each pair
of atoms $jk$, which can be written entirely in terms of model parameters
and amplitudes in modes with at most one atomic excitation:
\[
(2\epsilon_{r}+U_{jk})B_{rr}^{jk}+\Omega_{j}^{b}B_{er}^{jk}+\Omega_{k}^{b}B_{er}^{kj}+0=0,
\]
\[
\left(\Omega_{j}^{b}\right)^{*}B_{rr}^{jk}+(\epsilon_{e}+\epsilon_{r})B_{er}^{jk}+0+\Omega_{k}^{b}B_{ee}^{jk}=-D^{jk},
\]
\[
\left(\Omega_{k}^{b}\right)^{*}B_{rr}^{kj}+0+(\epsilon_{e}+\epsilon_{r})B_{er}^{kj}+\Omega_{j}^{b}B_{ee}^{kj}=-D^{kj},
\]
\[
0+\left(\Omega_{k}^{b}\right)^{*}B_{er}^{jk}+\left(\Omega_{j}^{b}\right)^{*}B_{er}^{kj}+2\epsilon_{e}B_{ee}^{jk}=-C^{jk},
\]
where we have defined,
\[
U_{jk}\equiv U(|x_{j}-x_{k}|),
\]
\[
C^{jk}\equiv\sum_{n}^{N_{cav}}(g_{jn}B_{ce}^{nk}+g_{kn}B_{ce}^{nj}),
\]
\[
D^{jk}\equiv\sum_{n}^{N_{cav}}g_{jn}B_{cr}^{nk}.
\]
These sets of four equations yield solutions of the form,
\[
B_{ee}^{nm}\equiv X_{ee}^{nm}C^{nm}+Y_{ee}^{nm}D^{nm}+Z_{ee}^{nm}D^{mn},
\]
\[
B_{er}^{nm}\equiv X_{er}^{nm}C^{nm}+Y_{er}^{nm}D^{nm}+Z_{er}^{nm}D^{mn},
\]
where $X_{ee}^{nm},$ $Y_{ee}^{nm},$ $Z_{ee}^{nm},$ $X_{er}^{nm},$
$Y_{er}^{nm},$ and $Z_{er}^{nm}$ are coefficients depending only
on model parameters. Substituting this solution into the remaining
$O(N)$ equations yields,
\begin{multline*}
C_{k}^{*}A_{e}^{m}=\sum_{p}^{N_{cav}}g_{mp}B_{cc}^{kp}+(\epsilon_{c}^{k}+\epsilon_{e})B_{ce}^{km}+\left(\Omega_{m}^{b}\right)^{*}B_{cr}^{km}+\\
\sum_{n\neq m}^{N_{at}}\sum_{p}^{N_{cav}}g_{nk}^{*}X_{ee}^{nm}g_{np}B_{ce}^{pm}+\sum_{n\neq m}^{N_{at}}\sum_{p}^{N_{cav}}g_{nk}^{*}X_{ee}^{nm}g_{mp}B_{ce}^{pn}+\\
\sum_{n\neq m}^{N_{at}}\sum_{p}^{N_{cav}}g_{nk}^{*}Y_{ee}^{nm}g_{np}B_{cr}^{pm}+\sum_{n\neq m}^{N_{at}}\sum_{p}^{N_{cav}}g_{nk}^{*}Z_{ee}^{nm}g_{mp}B_{cr}^{pn},
\end{multline*}
\textbf{
\begin{multline*}
C_{k}^{*}A_{r}^{m}=\Omega_{m}^{b}B_{ce}^{km}+\left(\epsilon_{c}^{k}+\epsilon_{r}\right)B_{cr}^{km}\\
+\sum_{n\neq m}^{N_{at}}\sum_{p}^{N_{cav}}g_{kn}^{*}X_{er}^{nm}g_{np}B_{ce}^{pm}+\sum_{n\neq m}^{N_{at}}\sum_{p}^{N_{cav}}g_{kn}^{*}X_{er}^{nm}g_{mp}B_{ce}^{pn}\\
+\sum_{n\neq m}^{N_{at}}\sum_{p}^{N_{cav}}g_{kn}^{*}Y_{er}^{nm}g_{np}B_{cr}^{pm}+\sum_{n\neq m}^{N_{at}}\sum_{p}^{N_{cav}}g_{kn}^{*}Z_{er}^{nm}g_{mp}B_{cr}^{pn},
\end{multline*}
}
\[
(\epsilon_{c}^{n}+\epsilon_{c}^{k})B_{cc}^{nk}+\sum_{m}^{N_{at}}g_{mn}^{*}B_{ce}^{km}+\sum_{m}^{N_{at}}g_{mk}^{*}B_{ce}^{nm}=C_{k}^{*}A_{c}^{n}\,\,\,\,\,(n\neq k),
\]
\[
2\epsilon_{c}^{k}B_{cc}^{kk}+\sqrt{2}\sum_{m}^{N_{at}}g_{mk}^{*}B_{ce}^{km}+0=\sqrt{2}C_{k}^{*}A_{c}^{k},
\]
which can be solved numerically. Finally, one obtains the amplitudes
for two photons occupying the cavity $B_{cc}^{nk}$, from which the correlation functions $G_{nk}$ can be calculated as, 
\[
\begin{cases}
G_{kk}=\frac{2|B_{kk}|^{2}}{|A_{k}|^{4}}\\
G_{nk}=\frac{|B_{kn}|^{2}}{|A_{k}|^{2}|A_{n}|^{2}} & n\neq k
\end{cases}.
\]


\end{document}